%% file: SUS-13-007_temp.tex
\pdfoutput=1

\documentclass[11pt,twoside,a4paper,cmspaper,final,collab]{cms-tdr}
\def\svnVersion{249998}\def\cmsCernNo{2013-209}\def\cmsCernDate{\today}\def\cmsMessage{Published in Physics Letters B as \href{http://dx.doi.org/10.1016/j.physletb.2014.04.023}{\doi{10.1016/j.physletb.2014.04.023.}}}

\begin{document}\cmsNoteHeader{SUS-13-007}

\hyphenation{had-ron-i-za-tion}
\hyphenation{cal-or-i-me-ter}
\hyphenation{de-vices}

\RCS$Revision: 238153 $
\RCS$HeadURL: svn+ssh://svn.cern.ch/reps/tdr2/papers/SUS-13-007/trunk/SUS-13-007.tex $
\RCS$Id: SUS-13-007.tex 238153 2014-04-25 07:57:49Z adamwo $
\providecommand{\ETslash}{\ensuremath{E_{\mathrm{T}}\hspace{-1.1em}/\kern0.45em}\xspace}
\providecommand{\met}{\ETslash}
\providecommand\stlep{\ensuremath{S_{\cmsSymbolFace{T}}^{\text{lep}}}\xspace}
\newcommand\Dphi{\ensuremath{\Delta\phi(\PW,\ell)}\xspace}
\newcommand\DP{\ensuremath{\Delta\phi}\xspace}
\newcommand\RCSx{\ensuremath{R_\mathrm{CS}}\xspace}
\newcommand\RCSpred{\ensuremath{R_\mathrm{CS}^\text{pred}}\xspace}
\newcommand\kCS{\ensuremath{\kappa_\mathrm{CS}}\xspace}

\newcommand{\njet}{\ensuremath{N_\text{j}}\xspace}
\newcommand{\nbjet}{\ensuremath{N_\cPqb}\xspace}
\newcommand{\Wjets}{\PW+jets\xspace}
\renewcommand{\chiz}{\ensuremath{\widetilde{\chi}^{0}_{1}}\xspace}

\newcommand\kLS{\ensuremath{\kappa_\mathrm{LS}}\xspace}
\newcommand\ktau{\ensuremath{\kappa_{\tau}}\xspace}
\renewcommand{\ETmiss}{\met}
\newcommand{\rThreeTwo}{\ensuremath{R_{32}}\xspace}

\newlength\cmsFigWidth
\ifthenelse{\boolean{cms@external}}{\setlength\cmsFigWidth{0.85\columnwidth}}{\setlength\cmsFigWidth{0.4\textwidth}}
\ifthenelse{\boolean{cms@external}}{\providecommand{\cmsLeft}{top}}{\providecommand{\cmsLeft}{left}}
\ifthenelse{\boolean{cms@external}}{\providecommand{\cmsRight}{bottom}}{\providecommand{\cmsRight}{right}}
\cmsNoteHeader{SUS-13-007} 
\title{Search for supersymmetry in pp collisions at $\sqrt{s}=8$\TeV in events with a single lepton, large jet multiplicity, and multiple b~jets}

\date{\today}

\abstract{
Results are reported from a search for supersymmetry
in $\Pp\Pp$ collisions at a center-of-mass energy of 8\TeV,
based on events with a single isolated lepton (electron or muon) and multiple jets, at least two of which are identified as b jets.
The data sample corresponds to an integrated luminosity of 19.3\fbinv recorded by the CMS experiment at the LHC in 2012.
The search is motivated by supersymmetric models that involve strong-production processes and cascade decays of new particles.
The resulting final states contain multiple jets as well as missing transverse momentum from weakly interacting particles.
The event yields, observed across several kinematic regions,
are consistent with the expectations from standard model processes.
The results are interpreted in the context of simplified supersymmetric scenarios with pair production of gluinos, where each gluino decays to a top quark-antiquark pair and the lightest neutralino.
For the case of decays via virtual top squarks, gluinos with a mass smaller than 1.26\TeV are excluded for low neutralino masses.
}

\hypersetup{%
pdfauthor={CMS Collaboration},%
pdftitle={Search for supersymmetry in pp collisions at sqrt(s) = 8 TeV in events with a single lepton, large jet multiplicity, and multiple b jets},%
pdfsubject={CMS},%
pdfkeywords={CMS, physics, supersymmetry, SUSY}}

\maketitle 
\section{Introduction}

This paper presents results from a search for new physics in
proton-proton collisions at a center-of-mass energy of 8\TeV
in events with a single lepton (electron or muon), missing transverse momentum, and multiple jets,
at least two of which are tagged as originating from bottom quarks (\cPqb-tagged jets).
This signature arises in models based on supersymmetry
(SUSY)~\cite{Wess:1974tw,Dimopoulos:1981zb,Nilles:1983ge,Haber:1984rc,Barbieri:1982eh,Dawson:1983fw},
which potentially offers natural solutions to limitations of the standard model (SM).
Large loop corrections to the Higgs boson mass could be cancelled by contributions from supersymmetric partners of SM particles.
Achieving these cancellations requires the gluino (\PSg) and top squark (\sTop), which are the SUSY partners of the gluon and top quark, respectively, to have masses less than about 1.5\TeV~\cite{Sakai:1981gr,Dimopoulos:1995mi,Papucci:2011wy,Brust:2011tb}.
Here and throughout this document we only consider the lighter of the two top squarks.
Extensive searches at LEP, the Tevatron, and the Large Hadron Collider (LHC) have not produced evidence for SUSY
(see Refs.~\cite{Chatrchyan:2013xna,SUS-12-010paper,SUS-11-028paper,Aad:2012naa,Aad:2012yr,Aad:2012ms,:2012ar} for recent results in the single-lepton topology).
For scenarios with mass-degenerate scalar partners of the first- and second-generation quarks, the mass limits generally lie well above 1\TeV.
However, viable scenarios remain with \sTop\ and \PSg\ masses below approximately 0.5 and 1.5\TeV, respectively.

In some of these scenarios top squarks are the lightest quark partners.
In R-parity conserving models \cite{Farrar:1978xj} this could lead to signatures with multiple \PW\ bosons, multiple \cPqb\ quarks, and two LSPs in the final state, where the LSP is the weakly interacting lightest SUSY particle.
The search described in this paper is designed to detect these signatures.
It focuses on gluino pair production, with subsequent gluino decay to two top quarks and the LSP (\chiz) through either a virtual or an on-shell top squark:
$\Pp\Pp \rightarrow \PSg\PSg$ with  $\PSg ( \rightarrow \sTop \cPaqt ) \rightarrow \cPqt\cPaqt\chiz$.
These decay chains result in events with high jet multiplicity, four \cPqb\ quarks in the final state, and large missing transverse momentum (\met).
The probability that exactly one of the four \PW~bosons decays leptonically is approximately 40\%, motivating a search in the single-lepton channel.

Three variations of this scenario, denoted as models A, B, and C, are considered in this analysis and implemented within the simplified model spectra (SMS) framework~\cite{ArkaniHamed:2007fw,Alwall:2008ag,Alves:2011wf}.
In model A (models B and C), gluinos are lighter (heavier) than top squarks and gluino decay proceeds through a virtual (real) \sTop.
For model A, the gluino and LSP masses $m_{\PSg}$ and $m_{\chiz}$ are allowed to vary.
For model B, we set $m_{\PSg}=1\TeV$ and vary $m_{\chiz}$ and the top squark mass $m_{\sTop}$.
For model C, $m_{\chiz}=50\GeV$ while $m_{\PSg}$ and $m_{\sTop}$ are varied.

The relevant backgrounds for this search arise from \ttbar, \Wjets, and single-top quark processes with small contributions from diboson,  {\ttbar}\cPZ, {\ttbar}\PW, {\ttbar}H, and Drell--Yan~(DY)+jets production.
The non-\ttbar backgrounds are strongly suppressed by requiring at least six jets, at least two of which are \cPqb-tagged.
The remaining background is dominated by \ttbar events with large \met, generated either by a single highly boosted \PW\ boson that decays leptonically (single-lepton event) or by two leptonically decaying \PW\ bosons (dilepton event).
Though \ttbar decays produce two true \cPqb\ quarks, additional \cPqb-tagged jets can arise because of gluon splitting to a \cPqb\cPaqb\ pair or from mistagging of charm-quark, light-quark, or gluon jets.

We search for an excess of events over SM expectations using two approaches.
The first approach is based on the distribution of \met in exclusive intervals of \HT, where \HT is the scalar sum of jet transverse momentum (\pt) values.
In this approach, we evaluate the \met distribution in the signal region of high \HT in two different ways~\cite{SUS-12-010paper,SUS-11-028paper}:
 by extrapolating from lower \HT and by using the charged-lepton momentum spectrum (this latter spectrum is highly correlated with the neutrino \pt spectrum in events with a leptonically decaying \PW\ boson and so carries information about \met).
In this approach combined signal regions for the \Pe\ and the $\mu$ channels are used since the differences in terms of backgrounds and sensitivity are small.

The second approach, which is new and described in more detail in this paper, is based on the azimuthal angle, \Dphi, between the reconstructed \PW-boson direction and the lepton.
The single-lepton background from \ttbar is suppressed by rejecting events with small \Dphi.
As will be shown later, this angle carries information similar to the transverse mass of the lepton and \met,
but has superior resolution.
The search is performed in different regions of the quantity \stlep,
defined as the scalar sum of the \met and lepton \pt.
The \stlep variable is a measure of the leptonic energy in the event and does not necessarily require high \met in order to be large.
SUSY events are expected to appear at large \stlep, where the contribution from SM processes is small.
The definition of signal regions in terms of \stlep allows the inclusion of events with lower \met than are included in the first approach.
Therefore the \Pe\ channel receives a small, but non-negligible correction for the presence of multijet events, and the \Pe\ and the $\mu$ channels are treated as separate signal regions.

The two approaches are complementary in the kinematic observables used and data samples exploited.
The first approach searches the tails of the single-lepton \ttbar-dominated sample at high \ETmiss and \HT with two independent methods, while the second approach uses \Dphi to reject that background process and search in a low-background region dominated by dilepton events in which one lepton is not identified or lies outside the acceptance of the analysis.
Together the two approaches provide a broad view of possible deviations from the standard model.

\section{Data sample and event selection}\label{sec:Selection}

The data used in this search were collected in proton-proton collisions at $\sqrt{s} = 8\TeV$ with the Compact Muon Solenoid (CMS) experiment in 2012 and correspond to an integrated luminosity of 19.3\fbinv.
The central feature of the CMS apparatus is a superconducting solenoid, providing a magnetic field of 3.8\unit{T}.
Within the superconducting solenoid volume are a silicon pixel and strip tracker, a lead tungstate crystal electromagnetic calorimeter, and a brass-scintillator hadron calorimeter.
Muons are measured in gas-ionization detectors embedded in the steel flux-return yoke outside the solenoid.
Extensive forward calorimetry complements the coverage provided by the barrel and endcap detectors.
The origin of the CMS coordinate system is the nominal interaction point.
The polar angle $\theta$ is measured from the counterclockwise beam direction and the azimuthal angle $\phi$ (in radians) is measured in the plane transverse to the beam axis.
The silicon tracker, the muon systems, and the barrel and endcap calorimeters cover the regions $\abs{\eta} < 2.5$, $\abs{\eta} < 2.4$, and $\abs{\eta} < 3.0$, respectively, where  $\eta = -\ln[\tan(\theta/2)]$ is the pseudorapidity.
A detailed description of the CMS detector can be found elsewhere~\cite{ref:CMS}.

Simulated event samples based on Monte Carlo (MC) event generators are used to validate and calibrate the background estimates from data and to evaluate the contributions for some small backgrounds.
The \MADGRAPH~\cite{madgraph}~5 generator with CTEQ6L1~\cite{Pumplin:2002vw} parton distribution functions (PDFs)
is used for \ttbar, \PW +jets, DY+jets, {\ttbar}\cPZ, {\ttbar}\PW, and multijet processes and the {\POWHEG}~1.0~\cite{powheg} generator for single-top-quark production.
The \PYTHIA~\cite{pythia}~6.4 generator is used to generate diboson and {\ttbar}H samples and to describe the showering and hadronization of all samples (the $\mathrm{Z2}^{*}$ tune~\cite{Chatrchyan:1363299} is used).
The cross sections used to scale the yields of these samples are calculated at next-to-leading (NLO) or higher order.
Decays of $\tau$ leptons are handled by \TAUOLA~\cite{tauola}.
The \GEANTfour~\cite{GEANT4} package is used to describe the detector response.

The SUSY signals for the three scenarios considered in this analysis are generated with \MADGRAPH and CTEQ6L1 PDFs.
In these scenarios, gluinos are pair-produced and decay into $\cPqt\cPaqt\chiz$, assuming the narrow-width approximation.
For the signal samples, the detector response is described using a fast simulation~\cite{Abdullin:2011zz}.
The fast simulation has been validated extensively against the detailed \GEANTfour simulation for the variables relevant for this search and efficiency corrections based on data are applied.
All simulated events are reweighted to match the multiplicity distribution of additional proton-proton collisions (``pileup'') as observed in data.

Events are selected online with either triple- or double-object triggers.
The triple-object triggers require a lepton with $\pt>15$\GeV, together with $\HT>350$\GeV and $\met > 45$\GeV.
The double-object triggers, which are used to select control samples and extend the \met acceptance in the approach based on \Dphi,  have the same \HT requirement, no $\met$ requirement, and a lepton \pt threshold of 40\GeV.
The trigger object efficiencies are measured in independently triggered control samples and found to reach a plateau at approximately 95\% for thresholds well below those used in the offline selection.
The measured trigger efficiencies are used to correct the simulation.

The preselection of events is based on the reconstruction of an  isolated lepton (\Pe~or $\mu$) and multiple jets and follows the procedure described in Ref.~\cite{SUS-12-010paper}.
Events are required to include at least one lepton with $\pt > 20\GeV$ and $\abs{\eta}<2.5$ (\Pe) or $\abs{\eta}<2.4$ ($\mu$).
Standard identification and isolation requirements~\cite{ref:eleID,CMS-PAPERS-MUO-10-004} are applied to reject backgrounds from
jets mimicking the lepton signature and from non-prompt leptons produced in semileptonic decays of hadrons within jets.
The isolation selection requires the sum of transverse momenta of particles in a cone of radius $\sqrt{(\Delta\eta)^2+ (\Delta\phi)^2}=0.3$ around the electron (muon) direction, divided by the \pt of the lepton itself, to be less than 0.15 (0.12).
The lepton efficiencies are measured with a ``tag-and-probe'' technique~\cite{CMS:2011aa} to be approximately 80\% for electrons and 95\% for muons.
The efficiencies vary by less than 20\% over the selected kinematic range and the average values agree to better than 1\% between data and simulation.

Jets are clustered from particles reconstructed with the particle-flow (PF) algorithm~\cite{ref:PAS-PFT-10-002},
which combines information from all components of the detector.
The clustering is performed with the anti-\kt clustering
algorithm~\cite{ref:antiKT} with a distance parameter of 0.5.
Jet candidates are required to satisfy quality criteria that suppress noise and spurious non-collision-related energy deposits.
Jets with $\pt > 40\GeV$ and $\abs{\eta}<2.4$ are considered in the analysis and are used to determine the number of selected jets \njet and $\HT$.
The missing transverse momentum is determined from the vector sum of the momenta of all particles reconstructed by the PF algorithm.
Jet and \met energies are corrected to compensate for shifts in the jet energy scale and the presence of particles from pileup interactions~\cite{CMS-PAPERS-JME-10-011}.

The number of \cPqb-tagged jets, \nbjet, is determined by applying the combined secondary vertex tagger~\cite{Chatrchyan:2012jua,CMS-PAS-BTV-13-001} to the selected jets.
At the working point used, this tagger has a roughly 70\% \cPqb-tag efficiency, and a mistag rate for light partons (charm quarks) of approximately 3\% (15--20\%).
Scale factors for the efficiencies and mistag rates relative to simulation are measured with control samples in data and applied in the analysis.

As the signal events are expected to exhibit a high level of hadronic activity and contain a large number of \cPqb\ quarks,
events are required to have $\HT > 400\GeV$.
In addition, at least two \cPqb-tagged jets and a total jet multiplicity $\njet \ge 6$ are required.
The SM background in this sample is dominated by \ttbar production.
Samples selected with the requirements $3\leq \njet \leq 5$ or $\nbjet<2$ are used to define background-dominated control regions.
Events with a second isolated lepton with $\pt > 15\GeV$ are vetoed
by the nominal signal selection to suppress contributions from dilepton \ttbar decays,
but such events are used as a control sample to measure the residual background from that process.

\section{Search in missing transverse momentum and \texorpdfstring{$\HT$}{HT}}\label{sec:METHT}

We now describe the background estimation method based on the evaluation of the \met spectrum. This method utilizes two techniques, as mentioned above, both of which were employed for previous CMS studies~\cite{SUS-12-010paper,SUS-11-028paper}.
The {\it lepton spectrum} (LS) method makes use of the similarity between the neutrino and charged-lepton \pt spectra in \PW~decays to predict the high-side tail of the \met distribution from single-lepton \ttbar decays based on the \pt distribution of charged leptons with high \pt~\cite{ref:leptonspectrum}.
The contributions from dilepton and $\tau$ lepton decays are predicted from single-lepton and dilepton control samples.
The {\it missing transverse momentum template} (MT) method uses a parametric description of the \met spectrum based on a fit to control regions at low \HT.
Through extensive use of data control samples, we avoid uncertainties related to potential mismodelling of SM yields by the simulation in the high \HT  and high \met tails.

We consider overlapping signal regions corresponding to lower limits for \HT ranging from 400 to 1000\GeV,
each of which provides sensitivity to a different SUSY-particle mass region.
The \met spectrum in these samples is divided into exclusive ranges: 150--250, 250--350, 350--450, and $>$450\GeV.
To increase the sensitivity, the search regions are further divided into events with $\nbjet = 2$ and ${\ge}3$.
The two background estimation methods provide direct predictions for events with two \cPqb-tagged jets.
The expected yields at higher \nbjet are obtained by extrapolating those predictions to the ${\ge}3$ \cPqb-jet case.

\subsection{Prediction of the single-lepton background for the LS method}\label{sec:LSsinglelepton}

The $\met$ spectrum of the single-lepton background is predicted with a method based on the similarity of the neutrino and charged lepton \pt spectra in \PW~decays.
In each event, the charged and neutral lepton \pt can be very different, but the distributions of the true neutrino \pt and the true lepton \pt are identical in the absence of \PW~polarization.
There are several effects that result in differences between the observed lepton and neutrino \pt spectra and for which corrections are derived:
\PW\ polarization,  the effect of a lepton \pt threshold, and the difference between the $\met$ and lepton-\pt resolutions.
The \PW-boson polarization in \ttbar decays is the dominant effect that causes a difference between the neutrino and lepton \pt spectra.
This polarization is well understood theoretically~\cite{ref:ttbarpolarization}
and accounted for in the simulation.
The difference between the $\met$ and lepton-\pt resolution is modeled with $\met$ resolution templates measured in multijet data samples.
These samples are used because the $\met$ resolution is dominated by the measurement of the hadronic activity in an event,
and these samples have little genuine $\met$.
These resolution templates are binned in \HT\ and \njet and are used to smear the lepton-\pt spectrum to account for the difference with respect to the $\met$ resolution.

The single-lepton $\met$ spectrum is predicted from the lepton \pt spectrum using the following steps.
First, the lepton \pt spectrum in a control sample, selected with lepton $\pt>50$\GeV and without a $\met$ requirement,
is smeared with the resolution templates.
The smeared distribution is then corrected with \pt-dependent scale factors \kLS to obtain a predicted $\met$ spectrum.
These scale factors are defined by $\kLS(\pt~\text{bin}) = N_{\rm true}(\met~\text{bin})/N_\text{pred}(\pt~\text{bin})$,
where $N_\text{pred}$ is the MC yield in a given bin of the smeared \pt distribution and
$N_\text{true}$ is the MC yield in the same bin of $\met$.
The latter includes only events with a single lepton at generator level,
while the former includes all events passing the selection of the control sample.
This definition ensures that the scale factors model the $\met$ distribution from only the single-lepton background without
contributions from $\tau$ leptons or dilepton backgrounds, which are predicted separately.
The calculation of the scale factor is dominated by the contributions of \ttbar events, but \PW+jets, DY+jets, single-top quark, {\ttbar}\cPZ,\ and {\ttbar}\PW\ events are included as well; the contribution of diboson events is negligible.
The impact of {\ttbar}\cPZ, {\ttbar}\PW, and {\ttbar}H events is insignificant, and for the results shown in this Section only the first  two categories have been included.
The dependence of the scale factor on lepton \pt primarily reflects the effect of the \PW-boson polarization in \ttbar decays.
The scale factor varies from around 1.0 for lepton \pt of 150 to 250\GeV, after $\met$ resolution smearing, to about 1.5 for $\pt > 450$\GeV.

Systematic uncertainties are evaluated by calculating the change induced in the
scale factors from various effects and propagating this change to the predicted yields.
The sources of systematic uncertainty are the jet and $\met$ scale, \PW~polarization in \ttbar decays and direct \PW\ production, \ttbar and
sub-dominant background cross sections, lepton efficiency, muon \pt scale, and DY+jets yield.
The dominant uncertainties arise from
the statistical uncertainties of the simulated samples used in the determination of the scale factor (9--49\%),
the jet and $\met$ scale (7--31\%, depending on \HT\ and $\met$),
and the \PW~polarization in \ttbar decays (2--4\%).

\subsection{Predictions of \texorpdfstring{$\tau$}{tau} lepton and dilepton backgrounds for the LS method}\label{sec:LStau}

Neutrinos from $\tau$\ lepton decays cause
the \ETslash and charged-lepton \pt spectra to differ.
Therefore, the SM background from $\tau$\ leptons is evaluated separately, following the procedure documented in Ref.~\cite{ref:leptonspectrum}.
While $\tau$-lepton decays are well simulated, their \pt spectra may not be.
Thus we apply $\tau$-lepton response functions derived from simulated \ttbar events to the \pt spectra of electrons and muons measured in single-lepton and dilepton control samples.
To suppress DY events in these control samples, same flavor dilepton events are rejected if they have $\ETslash<40$\GeV or a dilepton invariant mass within 20\GeV of $m_{\cPZ}$.
In these control samples, the \ETslash requirement is removed and a selection to reject DY events is applied.
The \HT and \njet requirements are loosened in the control sample used to estimate the background of events with hadronically decaying $\tau$ leptons.
For leptonic (hadronic) $\tau$-lepton decays, hereafter labelled $\tau_\ell$ ($\tau_\mathrm{h}$),
the response function is the distribution of the daughter lepton (jet) \pt as a fraction of the parent $\tau$ lepton \pt.
To predict the contribution to the \ETslash spectrum,
the observed lepton in the control sample is replaced by a lepton (or jet), with the transverse momentum sampled from the appropriate response
function; the difference between the sampled and original \pt is added
vectorially to the \ETslash.
This procedure is used to predict three background categories:
single $\tau_\ell$, $\ell + \tau_\mathrm{h}$, and $\ell + \tau_\ell$ events; the notation $\ell$ includes $\tau_{\ell}$ components.

The \ETslash spectrum obtained from applying the response functions to the control samples is corrected as a function of \ETslash and \HT for
branching fractions and efficiencies determined from MC simulation.
These correction factors are roughly 0.2, 0.9, and 0.6 for the
single $\tau_\ell$, $\ell + \tau_\ell$, and $\ell + \tau_\mathrm{h}$
backgrounds, respectively, in all \HT\ bins.
A correction is derived from simulation to account for a possible dependence on \met of the event selection and acceptance (note that this correction is consistent with one within the uncertainties).

SM backgrounds also arise from dilepton events. There are two categories of these events: those with both leptons reconstructed but where only one of the leptons is selected, and those with one lepton that is not reconstructed, which can occur either because of a reconstruction inefficiency or because the lepton lies outside the $\eta$ acceptance of the detector.
The estimate of the background from these processes is given by the simulated \met distribution, corrected by the ratio of the number of data to MC events in a dilepton control sample.
This sample is the same as that used in the $\ell + \tau_\ell$ background prediction,
but with an additional requirement of $\met>100$\GeV used to retain high trigger efficiency.
Systematic uncertainties for the dilepton background estimate arise from the uncertainty in
the data/MC scale factor, pileup, trigger and selection efficiencies,
and the top-quark \pt spectrum.

\begin{figure*}[tb!]
\centering
\includegraphics[width=.32\textwidth]{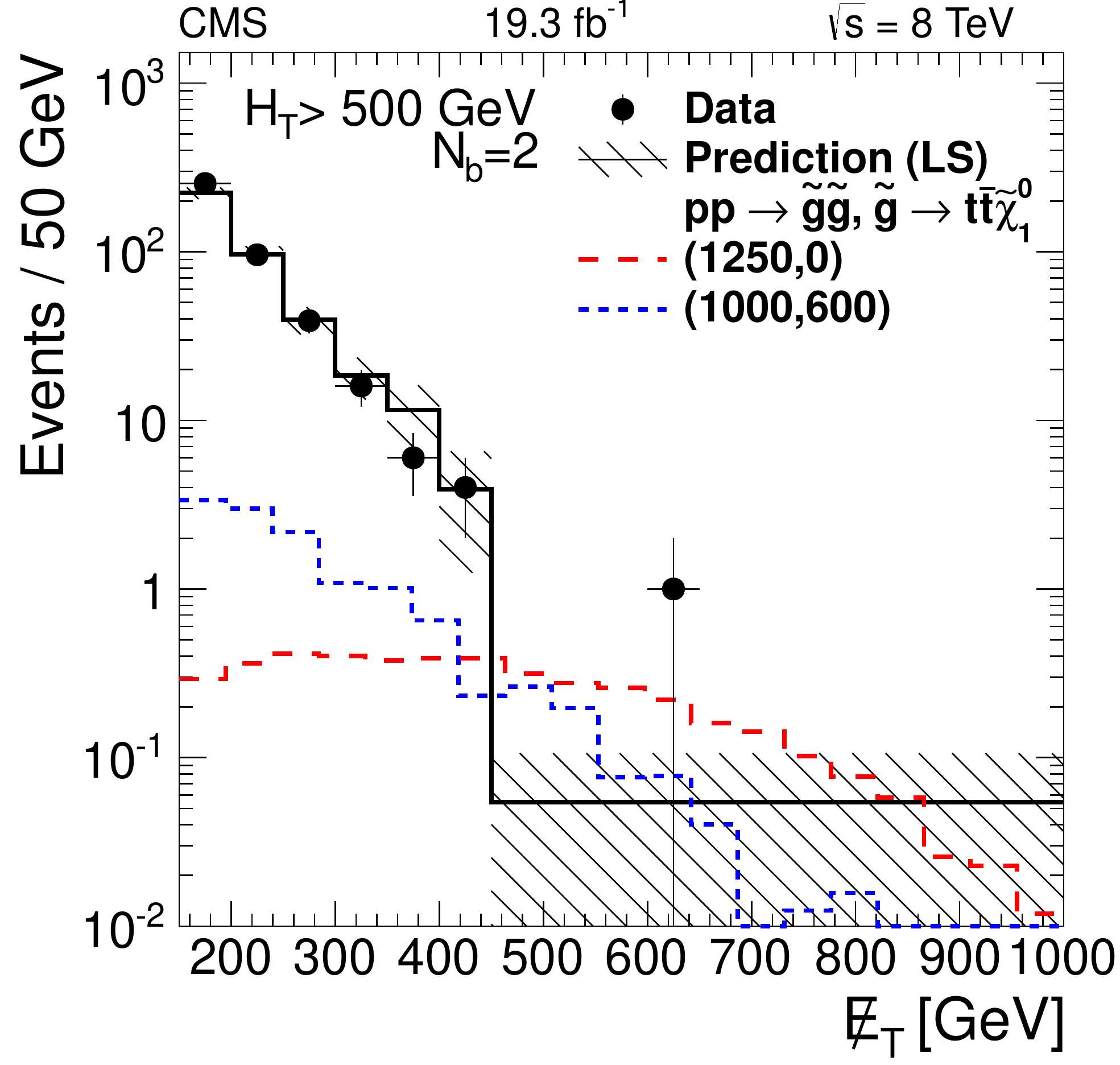}
\includegraphics[width=.32\textwidth]{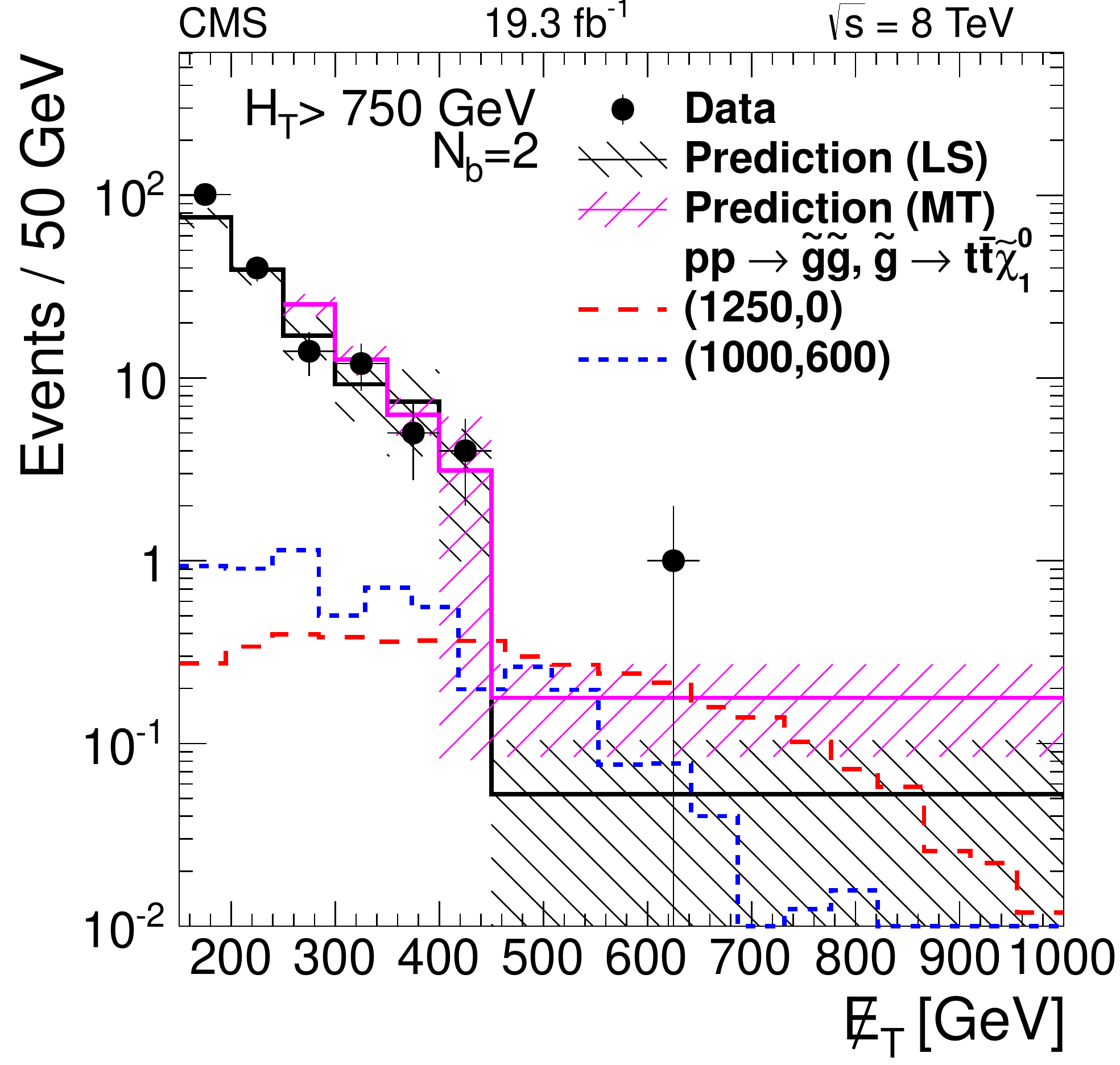}
\includegraphics[width=.32\textwidth]{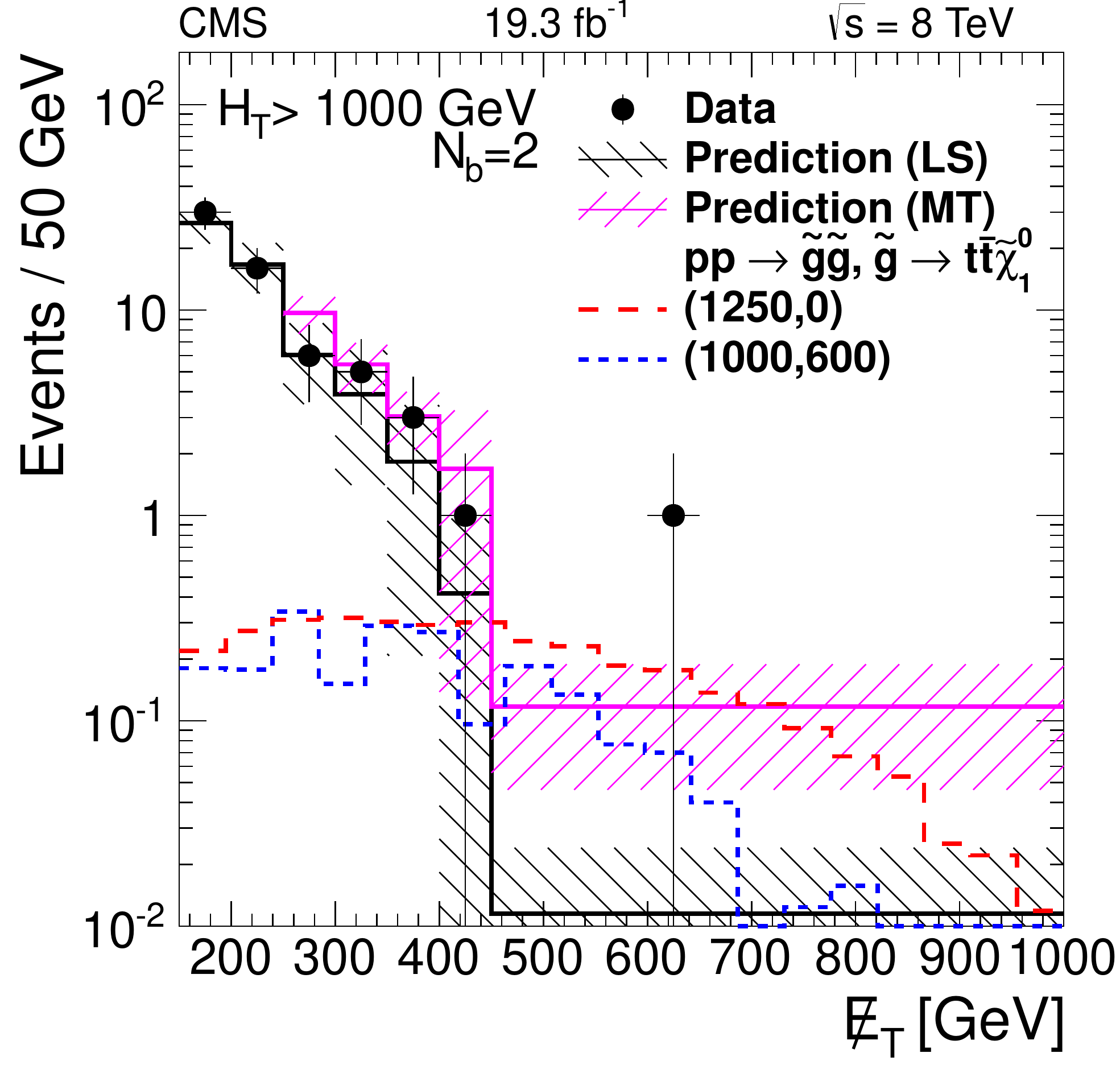}
\includegraphics[width=.32\textwidth]{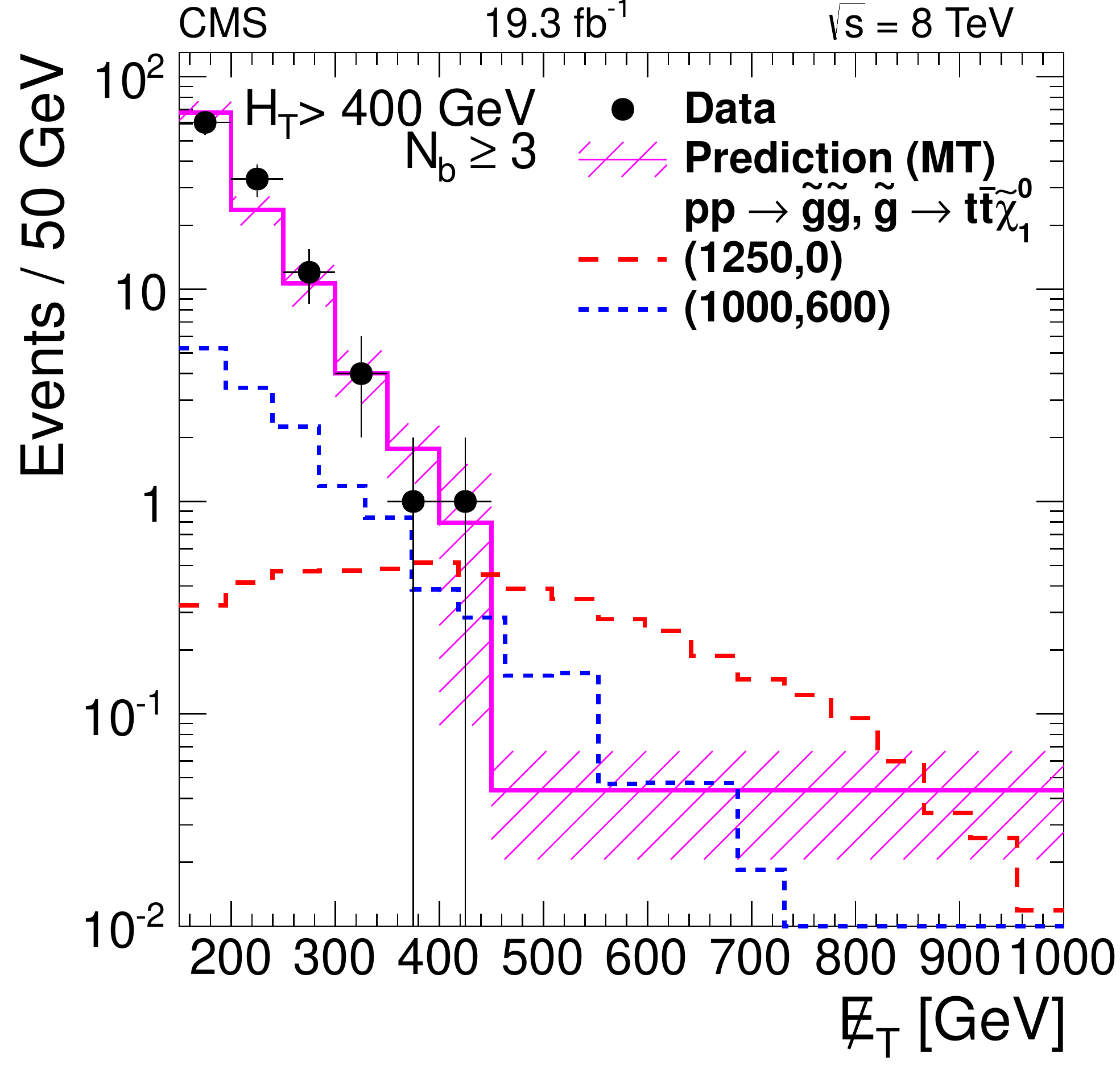}
\includegraphics[width=.32\textwidth]{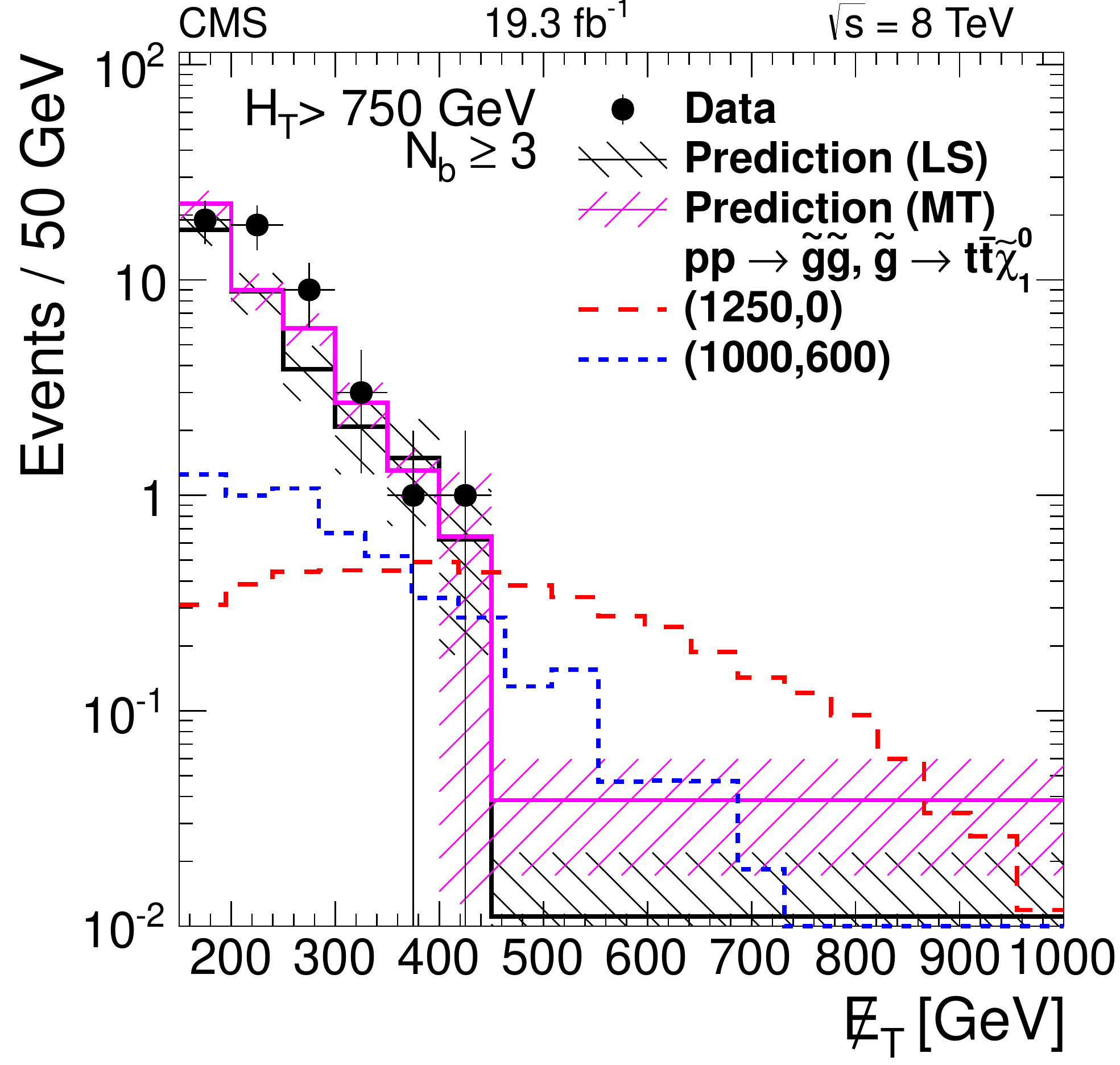}
\includegraphics[width=.32\textwidth]{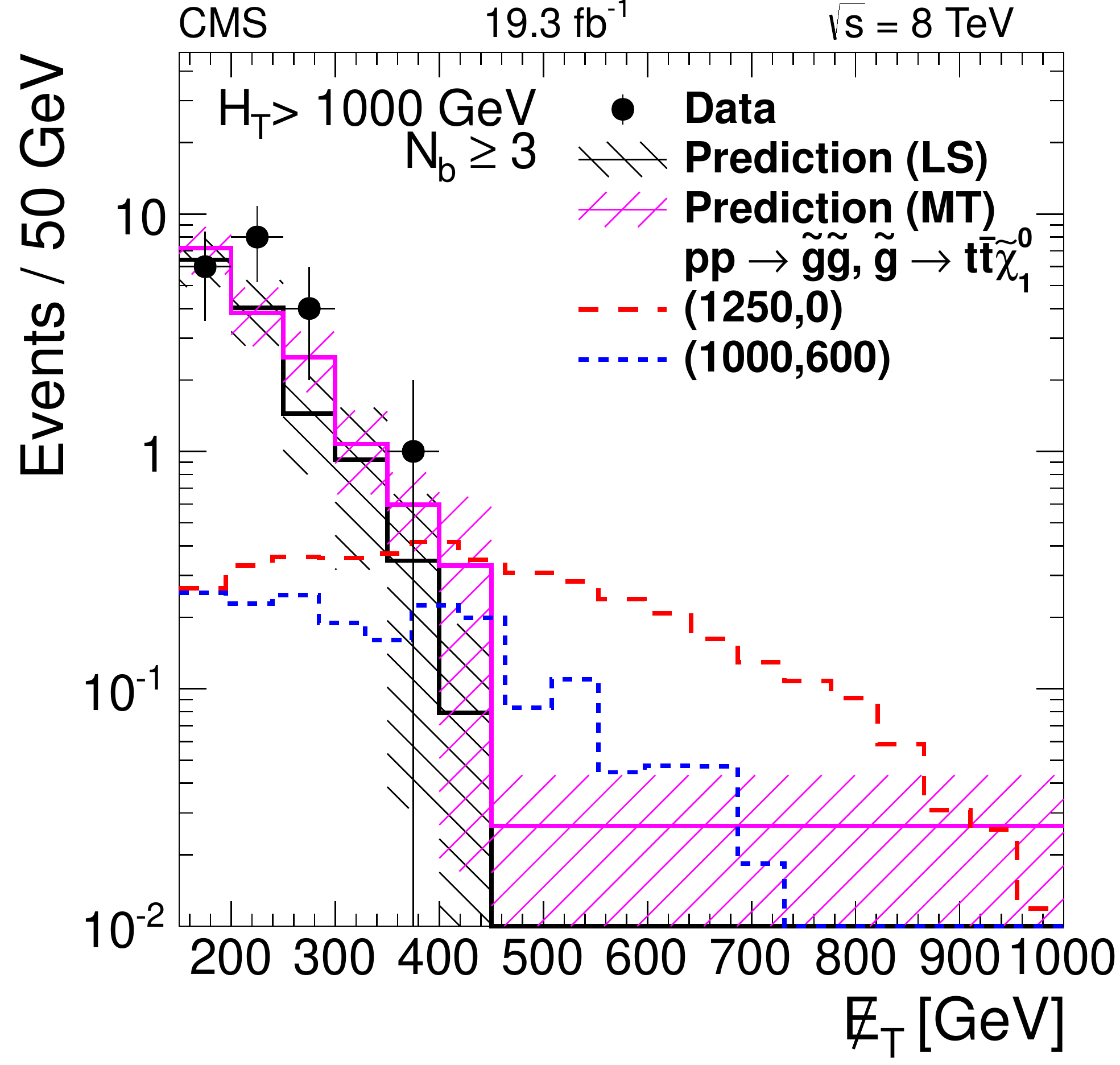}
\caption{
Observed \met distributions and the corresponding predictions from the LS and MT methods for the $\nbjet = 2$ (top) and ${\ge}3$ (bottom) bins.
The hatched areas show the combined statistical and systematic uncertainties of the predictions.
For purposes of comparison, the distributions for SUSY model A  with either $m_{\PSg}=1250\GeV$ and $m_{\chiz}=0\GeV$, or $m_{\PSg}=1000\GeV$ and $m_{\chiz}=600\GeV$, are shown.
Values and uncertainties for the prediction in the highest \ETmiss bin correspond to the average for the range 450--1000\GeV.
}\label{fig:resultsLSMT}
\end{figure*}

The background composition is similar in each of the LS signal regions.
For example, the signal region with $\HT > 500$~GeV, $\nbjet = 2$, and $350< \met <450$\GeV has predicted single-lepton and single-$\tau$ backgrounds of $11.6 \pm 5.2$ and $1.8 \pm 0.7$, respectively.
The remainder of the background prediction, consisting of $\ell + \tau_\mathrm{h}$, $\ell + \tau_\ell$, and dilepton events, is $2.0 \pm 1.1$.
The total yields for all signal regions are given in Table~\ref{tab:resultsLSMT} and Fig.~\ref{fig:resultsLSMT} shows the \met distributions.

\begin{table*}[tb!]
\topcaption{Observed yields in data and SM background predictions with their statistical and systematic uncertainties from the LS and MT methods.
For the MT method the low \met (150--250\GeV) and low \HT (400--750\GeV) regions in the $\nbjet = 2$ sample are used as control regions and are not shown in the table.
}\label{tab:resultsLSMT}
\centering
{\scriptsize
\begin{tabular}{c||c|lccc||c|lccc}
\hline
 \multirow{2}{*}{$\HT>400\GeV$}& \multicolumn{5}{c||}{ } & \multicolumn{5}{c}{$\nbjet \ge\ 3$} \\
 & \multicolumn{5}{c||}{ } & Obs. & & Pred. & $\pm$ stat. & $\pm$ syst. \\
\cline{1-1}\cline{7-11}
$150<\met<250\GeV$ &  \multicolumn{5}{c||}{ } & 94 & MT  & 92 & $\pm$  5 & $\pm$ 14 \\
$250<\met<350\GeV$ &  \multicolumn{5}{c||}{ } & 16 & MT  & 14.5 & $\pm$  1.3 & $\pm$  2.5 \\
$350<\met<450\GeV$ &  \multicolumn{5}{c||}{ } & 2 & MT  &  2.6 & $\pm$  0.4 & $\pm$  0.7 \\
$\met > 450\GeV$ &  \multicolumn{5}{c||}{ } & 0 & MT  &  0.8 & $\pm$  0.2 & $\pm$  0.4 \\
\hline
 \multirow{2}{*}{$\HT>500\GeV$} & \multicolumn{5}{c||}{$\nbjet = 2$} & \multicolumn{5}{c}{$\nbjet \ge\ 3$} \\
 & Obs. & & Pred. & $\pm$ stat. & $\pm$ syst. & Obs. & & Pred. & $\pm$ stat. & $\pm$ syst. \\ \hline
$150<\met<250\GeV$ & 350 & LS & 320 & $\pm$  16 & $\pm$  14 & 84 & LS & 71.1 & $\pm$  3.5 & $\pm$  8.3 \\
$250<\met<350\GeV$ & 55 & LS & 58.1 & $\pm$  7.2 & $\pm$  5.3 & 16 & LS & 12.4 & $\pm$  1.6 & $\pm$  1.8 \\
$350<\met<450\GeV$ & 10 & LS & 15.4 & $\pm$  4.3 & $\pm$  3.1 & 2 & LS & 3.1 & $\pm$  0.9 & $\pm$  0.7 \\
$\met > 450\GeV$ & 1 & LS &  0.7 & $^{+ 2.3}_{- 0.3}$ & $^{+ 2.0}_{- 0.2}$ & 0 & LS &  0.1 & $^{+ 0.5}_{- 0.0}$ & $^{+ 0.4}_{- 0.0}$ \\
\hline
 \multirow{2}{*}{$\HT>750\GeV$} & \multicolumn{5}{c||}{$\nbjet = 2$} & \multicolumn{5}{c}{$\nbjet \ge\ 3$} \\
 & Obs. & & Pred. & $\pm$ stat. & $\pm$ syst. & Obs. & & Pred. & $\pm$ stat. & $\pm$ syst. \\ \hline
\multirow{2}{*}{$150<\met<250\GeV$} & \multirow{2}{*}{141} & \multirow{2}{*}{LS} & \multirow{2}{*}{114.8} & \multirow{2}{*}{$\pm$ 9.4} & \multirow{2}{*}{$\pm$  6.9} & \multirow{2}{*}{37} & LS & 25.9 & $\pm$ 2.1 & $\pm$  3.1 \\
 & & & \multicolumn{3}{c||}{ } & & MT  & 31.8 & $\pm$  2.7 & $\pm$  4.8 \\
$250<\met<350\GeV$ & \multirow{2}{*}{26} & LS & 26.3 & $\pm$  4.9 & $\pm$  2.9 & \multirow{2}{*}{12} & LS & 5.9 & $\pm$  1.1 & $\pm$  1.0 \\
 & & MT  & 37.9 & $\pm$  4.0 & $\pm$  3.5 & & MT  &  8.5 & $\pm$  0.9 & $\pm$  1.6 \\
$350<\met<450\GeV$ & \multirow{2}{*}{9} & LS & 10.6 & $^{+ 3.8}_{- 3.7}$ & $\pm$  2.4 & \multirow{2}{*}{2} & LS & 2.1 & $\pm$  0.7 & $\pm$  0.5 \\
 & & MT  &  9.4 & $\pm$  1.4 & $\pm$  2.7 & & MT  &  1.9 & $\pm$  0.3 & $\pm$  0.6 \\
$\met > 450\GeV$ & \multirow{2}{*}{1} & LS &  0.6 & $^{+ 3.0}_{- 0.2}$ & $^{+ 1.9}_{- 0.2}$ & \multirow{2}{*}{0} & LS &  0.1 & $^{+ 0.7}_{- 0.0}$ & $^{+ 0.4}_{- 0.0}$ \\
 & & MT  &  3.1 & $\pm$  0.7 & $\pm$  1.5 & & MT  &  0.7 & $\pm$  0.2 & $\pm$  0.4 \\
\hline
 \multirow{2}{*}{$\HT>1000\GeV$} & \multicolumn{5}{c||}{$\nbjet = 2$} & \multicolumn{5}{c}{$\nbjet \ge\ 3$} \\
 & Obs. & & Pred. & $\pm$ stat. & $\pm$ syst. & Obs. & & Pred. & $\pm$ stat. & $\pm$ syst. \\ \hline
\multirow{2}{*}{$150<\met<250\GeV$} & \multirow{2}{*}{46} & \multirow{2}{*}{LS} &\multirow{2}{*}{43.2} & \multirow{2}{*}{$\pm$  6.1} & \multirow{2}{*}{$\pm$  3.7} & \multirow{2}{*}{14} & LS & 10.4 & $\pm$  1.5 & $\pm$  1.5 \\
 & & & \multicolumn{3}{c||}{ } & & MT  & 11.1 & $\pm$  1.6 & $\pm$  1.8 \\
\multirow{2}{*}{$250<\met<350\GeV$} & \multirow{2}{*}{11} & LS & 9.9 & $\pm$  3.1 & $\pm$  1.7 & \multirow{2}{*}{4} & LS & 2.4 & $\pm$  0.7 & $\pm$  0.5 \\
 & & MT  & 15.1 & $\pm$  2.5 & $\pm$  1.9 & & MT  &  3.6 & $\pm$  0.6 & $\pm$  0.8 \\
\multirow{2}{*}{$350<\met<450\GeV$} & \multirow{2}{*}{4} & LS &  2.2 & $^{+ 2.3}_{- 1.6}$ & $^{+ 2.2}_{- 0.7}$ & \multirow{2}{*}{1} & LS &  0.4 & $^{+ 0.5}_{- 0.3}$ & $^{+ 0.4}_{- 0.2}$ \\
 & & MT  &  4.7 & $\pm$  0.9 & $\pm$  1.5 & & MT  &  0.9 & $\pm$  0.2 & $\pm$  0.4 \\
\multirow{2}{*}{$\met > 450\GeV$} & \multirow{2}{*}{1} & LS &  0.1 & $^{+ 2.2}_{- 0.1}$ & $^{+ 3.5}_{- 0.1}$ & \multirow{2}{*}{0} & LS &  0.0 & $^{+ 0.4}_{- 0.0}$ & $^{+ 0.7}_{- 0.0}$ \\
 & & MT  &  2.0 & $\pm$  0.5 & $\pm$  1.1 & & MT  &  0.5 & $\pm$  0.1 & $\pm$  0.3 \\ \hline
\end{tabular}
}
\end{table*}

\subsection{The missing transverse momentum model in the MT method}\label{sec:MT}

For values of \met well above the \PW~boson mass, the SM \met distribution primarily arises from neutrino emission (genuine \met) and has an approximately exponential shape.
According to simulation, this distribution depends on \HT and, to a lesser extent, \njet and \nbjet, with only a small variation predicted for the non-exponential tails.
Empirically, we find that the genuine \met distribution from \ttbar events  (the leading background term) can be parametrized well with the Pareto distribution \cite{ref:Pickands}, which is widely used in extreme value theory:
\begin{equation}
f_\mathrm{P}(x;x_\text{min},\alpha,\beta) = \frac{1}{\alpha} \left( 1 + \frac{\beta ( x - x_\text{min} )}{\alpha} \right) ^{-\frac{1}{\beta}-1} ,
\label{eq:Pareto}
\end{equation}
where $x_\text{min}$, $\alpha$, and $\beta$ are the position, scale, and shape parameters, respectively.
Equation~\ref{eq:Pareto} yields an exponential function for $\beta=0$.
We set $x_\text{min}=150\GeV$, representing the lower bound of the \met spectrum to be described, while $\alpha$ and $\beta$ are determined from a fit to data.

Both the control regions used for a fit of the \met model to data and the signal regions have selection criteria applied to \HT.
Because of the correlation between the momentum of the leptonically decaying \PW\ boson and the momenta of the jets balancing it, restrictions on \HT affect the \met spectrum.
We describe the ratio between the \met spectrum after imposing a lower bound on \HT and the inclusive \met spectrum by a generalized error function (corresponding to a skewed Gaussian  distribution), similar to the approach described in Ref.~\cite{SUS-11-028paper}.
The evolutions with \HT of the location and variance parameters of this function are determined from simulation and found to be linear.
The results in simulation are found to be consistent with the dependence measured from a \ttbar-dominated control sample in data, defined by $\njet \ge 4$, $\nbjet \ge 2$, and $\HT > 400\GeV$.
Systematic uncertainties related to the error functions are determined from this comparison and from the difference between linear and quadratic models of the function parameters.
The \met spectrum in exclusive bins of $\nbjet$ is also affected by an acceptance effect due to the \pt requirement on the \cPqb-tagged jets:
in \ttbar events at low \HT, high values of \met correspond to low values of the \pt of the \cPqb\ quark associated with the leptonically decaying \PW\ boson and tend to move events to lower \cPqb-jet multiplicities.
We therefore apply an acceptance correction when applying the \met model to events with one or two \cPqb-tagged jets.
For $\nbjet=2$ and $150<\met<1000\GeV$, the size of the correction is 12\%  for $\HT=750\GeV$ and is smaller for larger \HT.

The \cPqb-jet multiplicity distribution is used to estimate the ratio of the \Wjets background to the \ttbar background  as a function of \HT.
The \HT distribution of \ttbar events is extracted from the $\nbjet = 2$ sample as described in Ref.~\cite{SUS-11-028paper}.
The contribution of \Wjets events for $\met>150\GeV$ is approximately~1\%.
Uncertainties related to other non-leading background components are estimated by varying the corresponding cross sections and are found to be small.
Based on the measured ratio of \Wjets to \ttbar background events, the Pareto distribution describing the leading background term is combined with the shape of the \Wjets \met distribution from simulation to form the full model describing the genuine  \met distribution of SM events.

\subsection{The fit to the missing transverse momentum spectrum in the MT method}\label{sec:MTfit}

The model for genuine \met in SM events is convolved with the \met resolution templates described in Section~\ref{sec:LSsinglelepton} and used in a simultaneous fit to the \ETmiss shapes in control regions in the $\nbjet = 1$ and $\nbjet = 2$ bins.
The control regions are chosen in order to ensure reasonably small statistical uncertainties and to limit potential contributions from signal events:
for events with two \cPqb-tagged jets the control region is defined by $400 < \HT < 750\GeV$ and $150 < \ETmiss < 400\GeV$,
while for one \cPqb-tagged jet it is extended to $400<\HT<2500\GeV$ and $150 < \ETmiss < 1500\GeV$.

Because of limited statistical precision in the control regions,
we are unable to obtain a reliable estimate of $\beta$ from data.
We use a constraint from simulation together with an uncertainty derived from a comparison between data and simulation in control regions with lower jet multiplicity.
The constraint is implemented as a Gaussian term corresponding to the value and its statistical uncertainty obtained from simulation, $\beta = 0.03 \pm 0.01$.
The prediction from simulation for $\njet = 3$--5 is $\beta = 0.15$--0.05, consistent with the data.
The maximum difference between data and simulation in any of these three \njet bins of 0.05 is used to define a systematic uncertainty in the prediction.
The parameters of the error function (Section~\ref{sec:MT}) are constrained by Gaussian terms reflecting the respective values and covariance from simulation.

The predictions for the $\nbjet=2$ signal regions are obtained by integrating the function representing the \ETmiss model over the relevant \ETmiss range and summing over the \HT bins.
In each \HT bin, the predicted distribution is scaled to match the observed number of events in the normalization region defined by $150 < \ETmiss < 250\GeV$.
The statistical uncertainties of the predictions are evaluated by repeating the procedure using parameter values randomly  generated according to the results of the fit, including   the covariance matrix.
The predictions are stable to within 1\% if the \met model described in Ref.~\cite{SUS-11-028paper} is  used in place of the model described here.

The results of the MT method can be affected by several systematic uncertainties that are related to detector effects, assumptions made on the shape of the distribution, as well as theoretical uncertainties and the contamination due to non-leading backgrounds.
Systematic uncertainties related to the  jet and \ETmiss scale, lepton reconstruction efficiencies, \PW-boson polarization in \ttbar events, and cross sections of non-leading backgrounds are evaluated in the same way as for the LS method (Section~\ref{sec:LSsinglelepton}).
Effects due to \cPqb-jet identification efficiencies and pileup are also taken into account.
In addition, the following uncertainties specific to the MT method are considered.
The $\beta$ parameter and parameters of the error function are  varied as described above.
The differences with respect to the  standard result define the systematic uncertainty for each signal  region.
The effects of a possible residual non-linearity in the error function parameters versus \HT are also taken into account.
To test the validity of the method, the procedure is applied  to simulated events.
The resulting background predictions are  found to be statistically consistent with the true numbers from simulation.
Conservatively, the maximum of the relative difference and its uncertainty are assigned as a further systematic uncertainty (``closure'').
The dominant contributions to the systematic uncertainty are related to the $\met$ model (1--35\%, depending on the \HT and \met bin) and the closure (8--43\%).
Uncertainties related to the theoretical predictions of the cross section for SM backgrounds, the jet and \met scale, and pileup contribute each with less than 5\%.

\subsection{Background estimation in the \texorpdfstring{$\nbjet \ge3$}{Nb >=3} bin}\label{sec:MTthreetags}

The numbers of data events in the $\nbjet \ge 3$ control samples are too low for an application of the LS or MT technique.
Therefore we estimate the background for high \cPqb-jet multiplicities by applying to the background predictions for $\nbjet = 2$ transfer factors
(\rThreeTwo) that give the ratio of the number of events with $\geq 3$ and $= 2$ \cPqb-tagged jets for each of the signal regions.
The central values for the \rThreeTwo factors are determined from simulation.
The scale factors \rThreeTwo increase with jet multiplicity from approximately~0.05 for events with three jets to approximately~0.2 in events with $\geq 6$ jets because of the higher probability of misidentifying one or more jets.
For constant jet multiplicity they do not demonstrate a strong dependence on \HT.

The ratios between $\nbjet \ge 3$ and $= 2$ events in data and simulation could differ because of incorrect modeling of the heavy-flavor content, the jet kinematics, and uncertainties in the \cPqb-tagged jet misidentification rates.
To probe the impact of the first source of uncertainties, the weight of events with at least one \cPqc\ quark is varied by ${\pm}50\%$.
A variation of the same size is applied to events with additional \cPqb- or \cPqc-quark pairs.
The effect of possible differences between data and simulation in the kinematics of the system of non-\cPqb\ jets on \rThreeTwo is tested in a control sample with exactly two \cPqb-tagged jets.
The remaining jets in the event are randomly assigned a parton flavor: one jet is marked as a \cPqc-quark jet, while the others are marked as light-quark jets.
Based on this assignment the ratio of probabilities to tag at least one additional jet is calculated.
This procedure is applied to both data and simulation.
Good agreement is found and the residual difference is interpreted as a systematic uncertainty.
The uncertainty related to \cPqb-tagged jet misidentification is evaluated from the uncertainties of the misidentification scale factors relative to simulation.
The total systematic uncertainties for \rThreeTwo are approximately~9--19\% depending on the signal region.
An additional verification is performed in two control regions at higher lepton \pt (${>}30\GeV$), lower \HT (${<}400\GeV$) and \met ($150<\met<250\GeV$), and $\njet = 5$ or ${\ge}6$.
In both regions the values of \rThreeTwo obtained in simulation are compatible with the ones observed in data.

In the LS method, the transfer factors are applied to the signal regions for $\HT>500$, $750$, and $1000\GeV$.
In the MT method, signal regions for $\HT>400\GeV$ and $150<\met<250\GeV$ are added for the $\nbjet \ge 3$ bin, corresponding to the limits of the control and normalization regions in the $\nbjet = 2$ bin, respectively.

\subsection{Results for signal regions in missing transverse momentum and \texorpdfstring{\HT}{HT} bins}\label{sec:resultsLSMT}

The predictions of both methods are compared with the observed number of events in Table~\ref{tab:resultsLSMT}.
For the LS method the predictions consist of the single-lepton and $\tau$-lepton backgrounds with a small contribution from dilepton events.
Drell-Yan events are heavily suppressed by the $\njet$, $\nbjet$, and kinematic requirements.
The yield of this small component of the background is taken from simulation.
For the MT method the predictions consist of the inclusive estimation of the leading backgrounds.
Additional contributions to the signal regions from multijet events are heavily suppressed, but their cross section is large and not precisely known.
Therefore, they are predicted from data based on scaling the sideband of the relative lepton isolation distribution.
These contributions are neglected as they are found to constitute 1\% or less of the total background in all cases.

The corresponding observed and predicted \ETmiss spectra are shown in Fig.~\ref{fig:resultsLSMT} for the two \cPqb-jet multiplicity bins and different \HT requirements.
The two methods differ in their leading systematic terms  and in the correlations they exhibit between the background  predictions in different signal regions.
The predictions are consistent, an indication of the robustness of the methods.
No excess is observed in the tails of the \met distributions with respect to the expectations from SM processes.
The results are interpreted in terms of upper limits on the production cross section for different benchmark models in Section~\ref{sec:Interpretation}.

\section{Search using \texorpdfstring{$\stlep$}{STlep} and \texorpdfstring{\Dphi}{DPhi}}\label{sec:DPhi}

After applying the selection criteria in Section~\ref{sec:Selection}, the sample is  dominated by single-lepton \ttbar events.
In the {\it delta phi} (\DP) analysis method, this background is further reduced by applying a requirement on the azimuthal angle between the \PW-boson candidate and the charged lepton.
The \PW-boson candidate transverse momentum is obtained as the vector sum of the lepton \pt and the \met vectors.
For single-lepton \ttbar events, the angle between the \PW-boson direction and the charged lepton has a maximum value, which is fixed by the mass of the \PW\ boson and its momentum.
Furthermore, the requirement (direct or indirect) of large \met selects events in which the \PW\  boson yielding the lepton and the neutrino is boosted, thus resulting in a fairly narrow distribution in $\Delta\phi(\PW,\ell)$.
On the other hand, in SUSY decays, the ``effective \PW\  boson'' that is formed from the vector sum of the transverse momenta of the charged lepton and the \met vector will have no such maximum.
Since the \met results mostly from two neutralinos, the directions of which are
largely independent of the lepton flight direction, the $\Delta\phi(\PW,\ell)$ distribution is expected to be flat.

\begin{figure*}[tb!hp]
 \centering
  \includegraphics[width=0.31\linewidth]{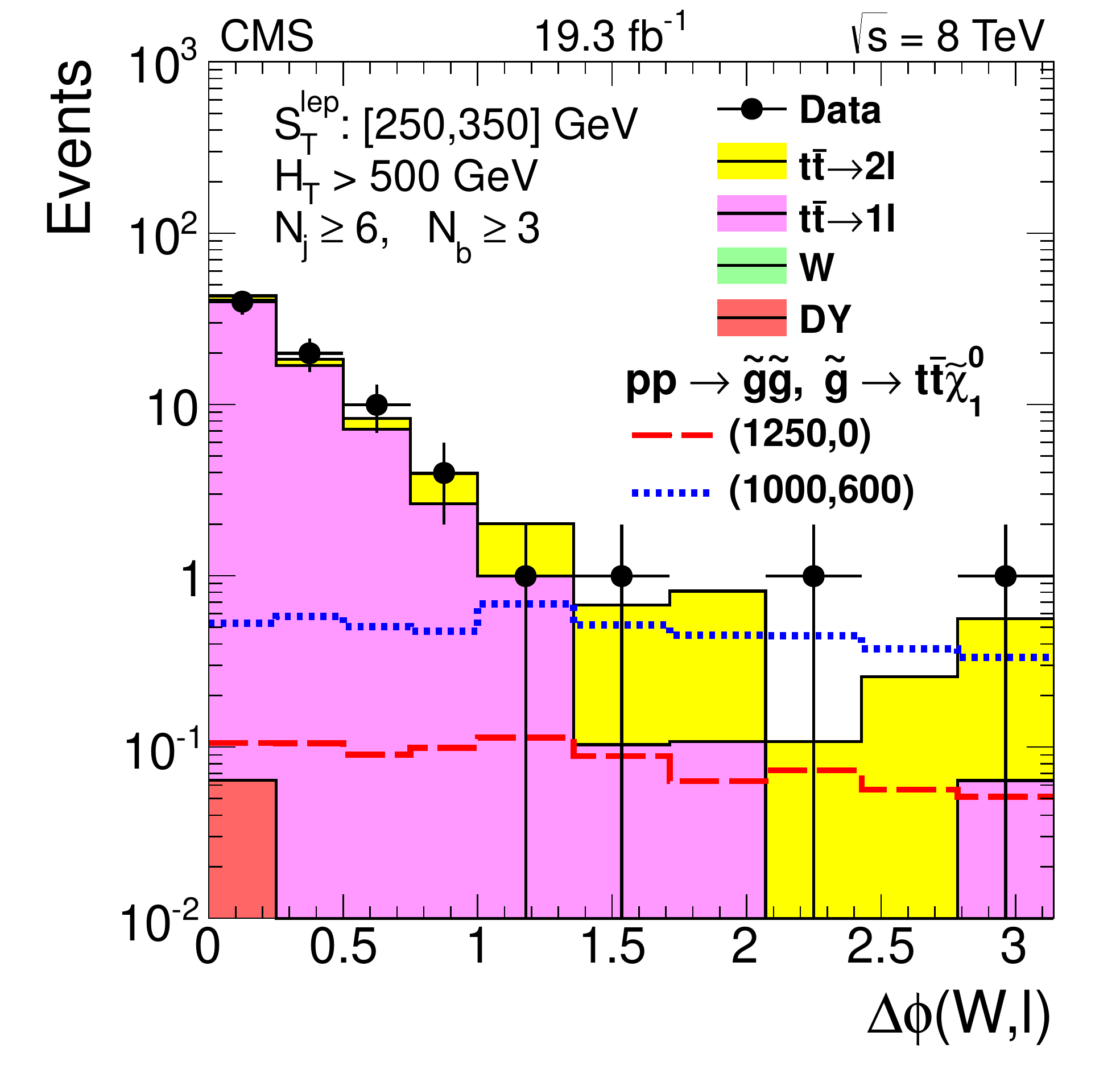}
  \includegraphics[width=0.31\linewidth]{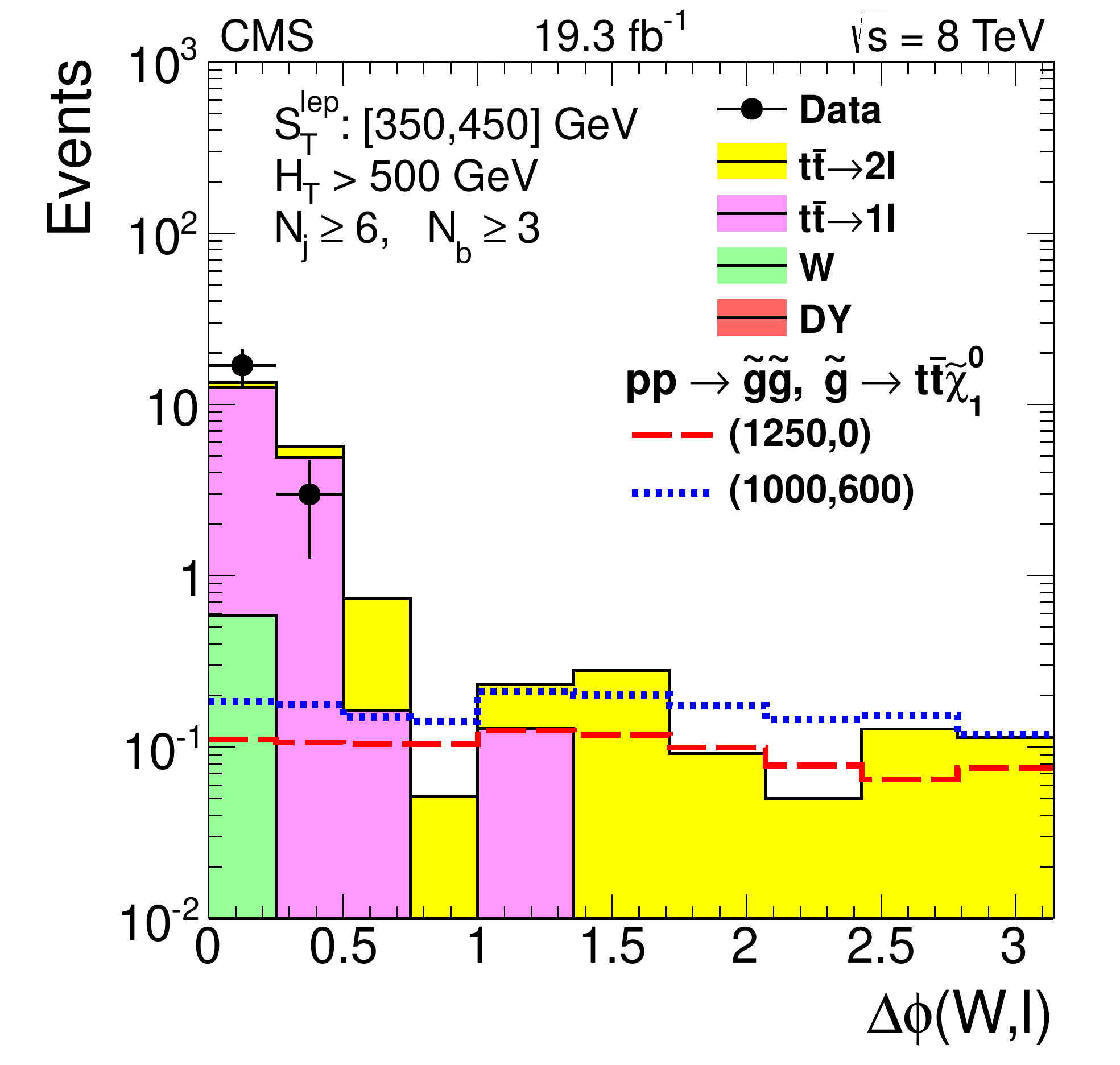}
 \includegraphics[width=0.31\linewidth]{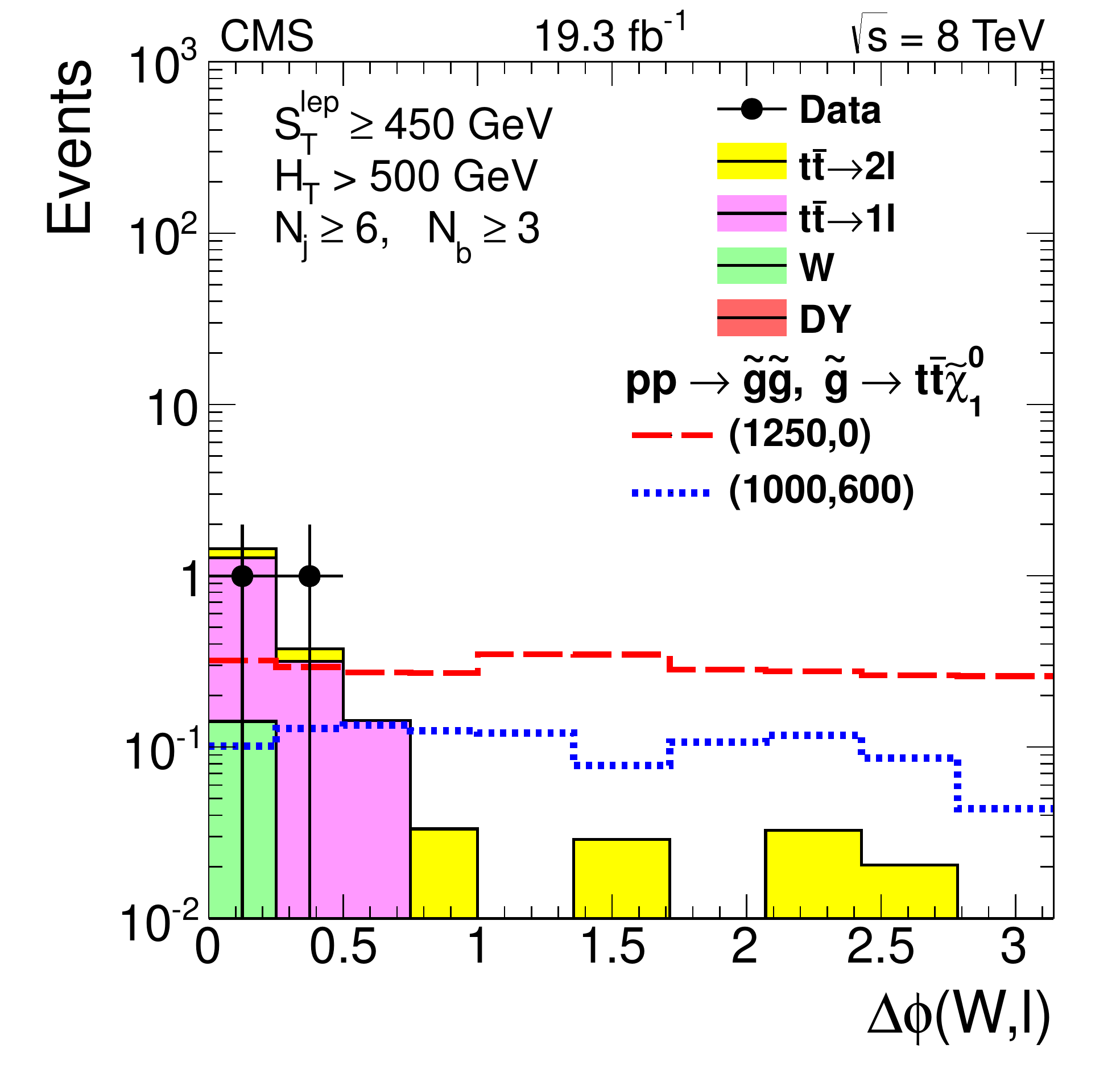}
 \caption{
The $\Delta\phi(\PW,\ell)$ distribution in simulation and data for the
combined \Pe\ and $\mu$ channels with $\nbjet \ge 3$ and $\njet \geq
6$. The SM simulation is normalized to the data in the control region
($\Delta\phi(\PW,\ell)<1$). The simulated SM yields in the signal region
($\Delta\phi(\PW,\ell)>1$) are shown only for illustration, as the
actual estimate is obtained with the procedure described in the text.
The distributions expected for signal are illustrated using two mass points from model A, with masses specified as $(m_{\PSg},m_{\chiz})$ in\GeV.
Left: $250 < \stlep < 350\GeV$, center: $350 < \stlep < 450\GeV$, and right: $\stlep > 450\GeV$.
}\label{Dphi_MC_muon}
\end{figure*}

Distributions of $\Delta\phi(\PW,\ell)$ in different \stlep bins are shown for the $\nbjet\ge 3$ and $\njet\geq 6$ samples in Fig.~\ref{Dphi_MC_muon}.  We select $\Delta\phi(\PW,\ell)>1$ as the signal region.
The complementary sample, events with $\Delta\phi(\PW,\ell)<1$, constitutes the control region.
It can be seen that this selection is effective in reducing the background from single-lepton \ttbar decays; the dominant background in the signal regions comes from dilepton \ttbar events.
Table~\ref{Tab:default_muon} shows the event yields from simulation for the signal and control regions in different \stlep bins for events with $\nbjet \ge 3$, which have the highest sensitivity to the SUSY signal.
The contributions from {\ttbar}\cPZ, {\ttbar}\PW, {\ttbar}H, and diboson events are insignificant and have not been used for the results shown in this Section.
The search also uses events with $\nbjet = 2$, albeit with smaller sensitivity.

\begin{table*}[b!ht]
\topcaption{Event yields for the combined \Pe\ and $\mu$ channels, as predicted by simulation, for $\njet\ge 6$ and $\nbjet\ge 3$.
The \RCSx column lists the ratio of yields in the signal and control regions.
The yields for signal benchmark points are shown for comparison, with the (\PSg,\chiz) masses (in \GeVns) listed in brackets. The uncertainties are statistical only.}
\label{Tab:default_muon}
\centering
\resizebox{\textwidth}{!}
{
\begin{tabular}{c||c|c|c||c|c|c||c|c|c} \hline
		& \multicolumn{3}{c||}{$250<\stlep<350\GeV$ } &  \multicolumn{3}{c||}{$350<\stlep<450\GeV$ } & \multicolumn{3}{c}{$\stlep >450\GeV$} \\ \hline
  Sample            & Signal & Control & \RCSx & Signal & Control &
  \RCSx & Signal & Control & \RCSx\\ \hline \hline
\ttbar$(1\ell)$                  & 0.8 $\pm$ 0.2 & 43.2 $\pm$ 1.8 &
  0.02 & 0.1 $\pm$ 0.1 & 11.6 $\pm$ 1.0 & 0.01 & $<$0.01 & 3.4
    $\pm$ 0.5 & n/a \\
\ttbar$(\ell\ell)$                & 2.0 $\pm$ 0.3 & 4.0 $\pm$ 0.4 &
    0.51
    & 0.5 $\pm$ 0.1 & 1.6 $\pm$ 0.3 & 0.34 & 0.2 $\pm$ 0.1 & 0.6
    $\pm$ 0.2 & 0.35 \\
\PW                                 & $<$0.23 &  $<$0.23  & n/a &
    $<$0.24
    & 0.4 $\pm$ 0.4 & n/a & $<$0.22 & 0.3 $\pm$ 0.3 & n/a \\

DY                                 & $<$0.03 & $<$0.03  & n/a
    & $<$0.02 & $<$0.02 & n/a & $<$0.03 & $<$0.03 & n/a \\

Multijet                               &  $<$0.05 & $<$ 0.05 & n/a  & $<$
    0.01 & $<$ 0.01 & n/a  & $<$ 0.01 & $<$ 0.01  & n/a   \\

Single \cPqt                      & 0.4 $\pm$ 0.2 & 1.9 $\pm$ 0.3 &
0.21 & $<$0.08 & 0.8 $\pm$ 0.2 & n/a & $<$0.08 & 0.4 $\pm$ 0.2 & 0.04 \\
\hline
SM all                            & 3.3 $\pm$ 0.4 & 49.0 $\pm$ 1.8 &
    0.07 & 0.6 $\pm$ 0.2 & 14.4 $\pm$ 1.1 & 0.04 & 0.2 $\pm$ 0.1
    & 4.7 $\pm$ 0.7 & 0.05 \\ \hline \hline
  Model A            & Signal & Control & \RCSx & Signal & Control &
  \RCSx & Signal & Control & \RCSx\\ \hline \hline
 (1000,600)        			  			
& 2.80 $\pm$ 0.10 & 2.09 $\pm$ 0.09 & 1.34
& 1.00 $\pm$ 0.06 & 0.65 $\pm$ 0.05 & 1.54
& 0.55 $\pm$ 0.05 & 0.49 $\pm$ 0.04 & 1.13 \\ \hline
 (1250,0)                   											
& 0.45 $\pm$ 0.01 & 0.40 $\pm$ 0.01 & 1.12
& 0.56 $\pm$ 0.02 & 0.42 $\pm$ 0.01 & 1.32
& 1.78 $\pm$ 0.03 & 1.16 $\pm$ 0.02 & 1.54 \\ \hline
\end{tabular}
}
\end{table*}

\subsection{Prediction of standard model background}

\begin{figure*}[tb!]
 \centering
  \includegraphics[width=0.32\linewidth]{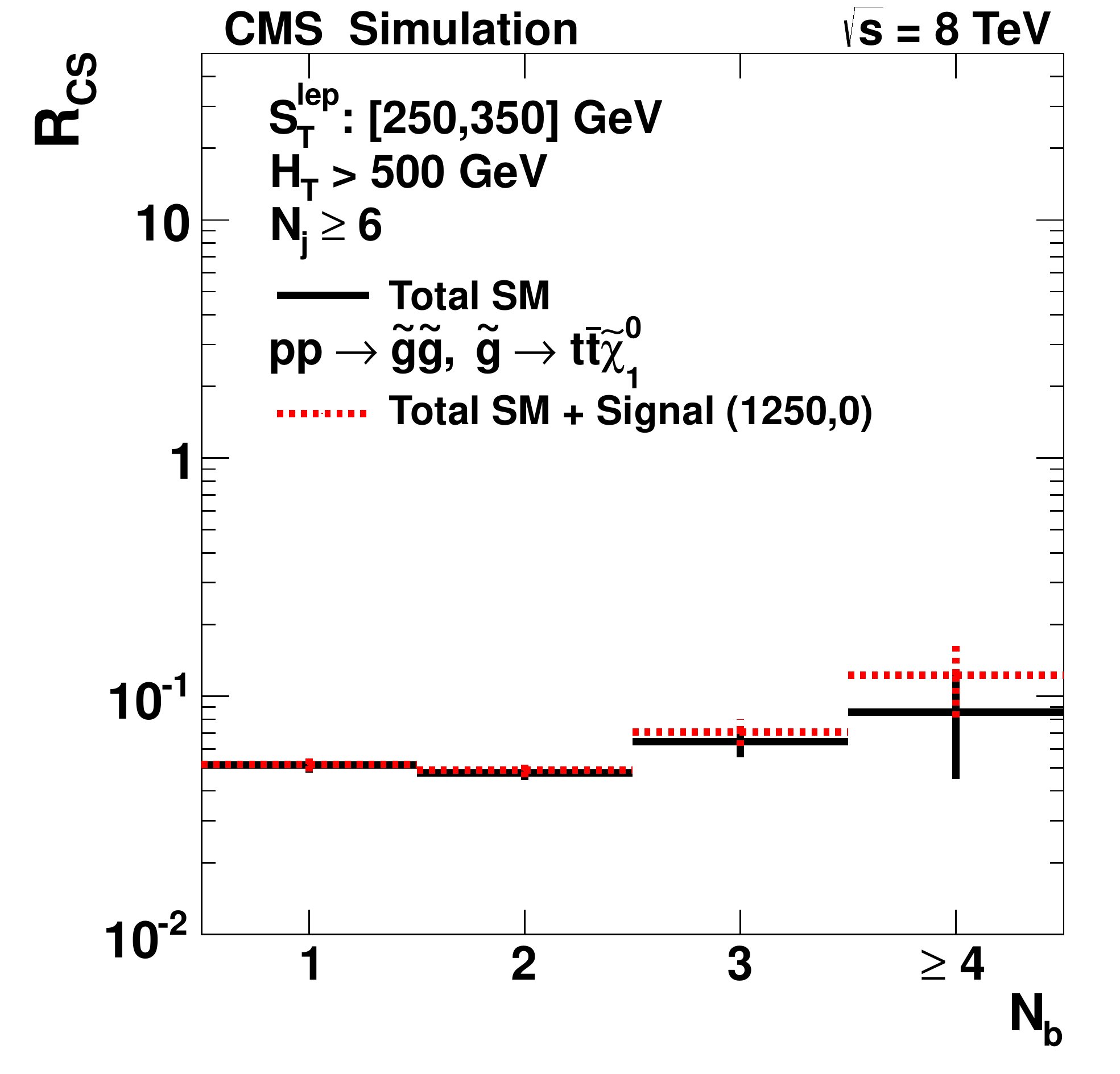}
  \includegraphics[width=0.32\linewidth]{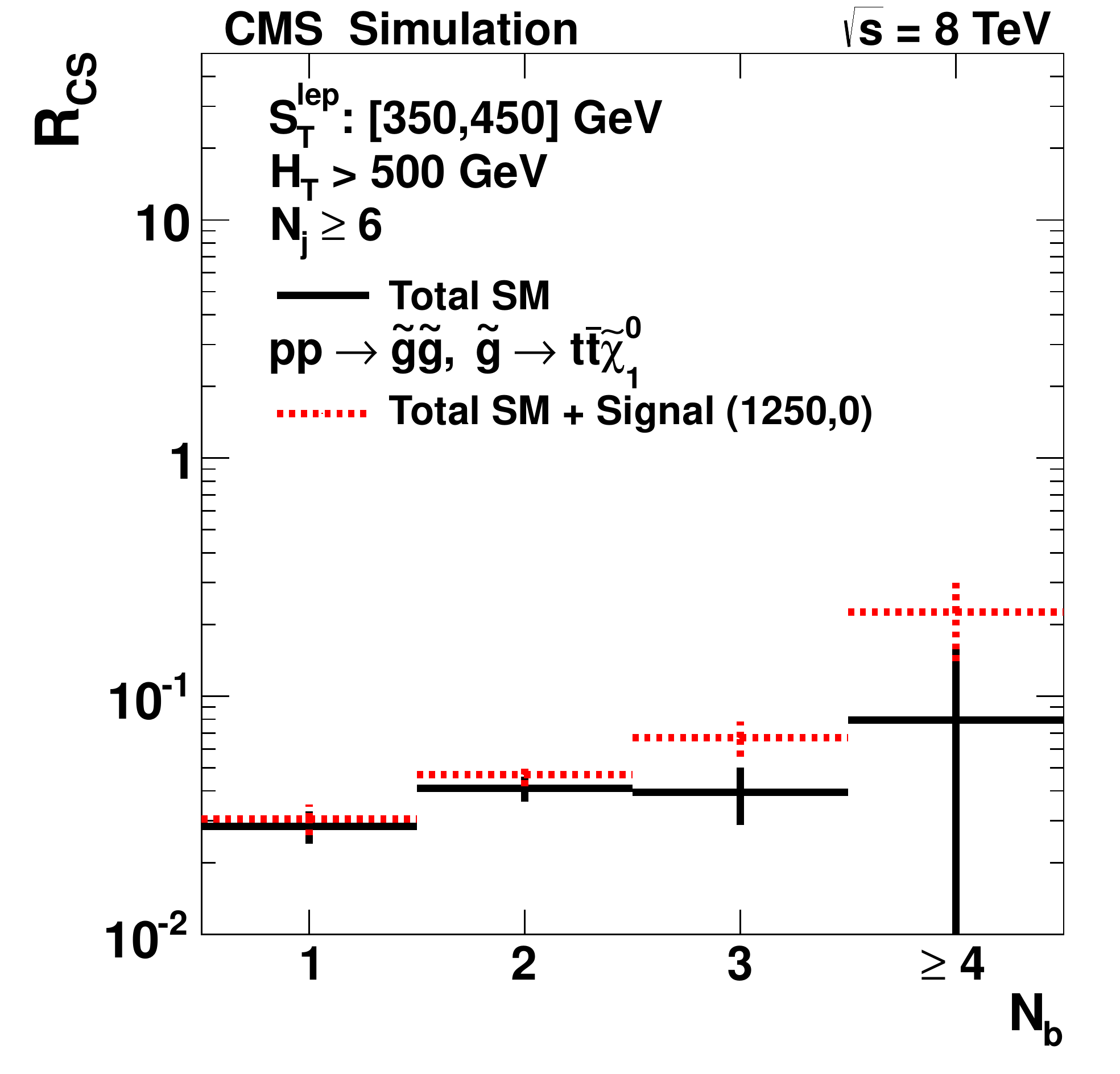}
  \includegraphics[width=0.32\linewidth]{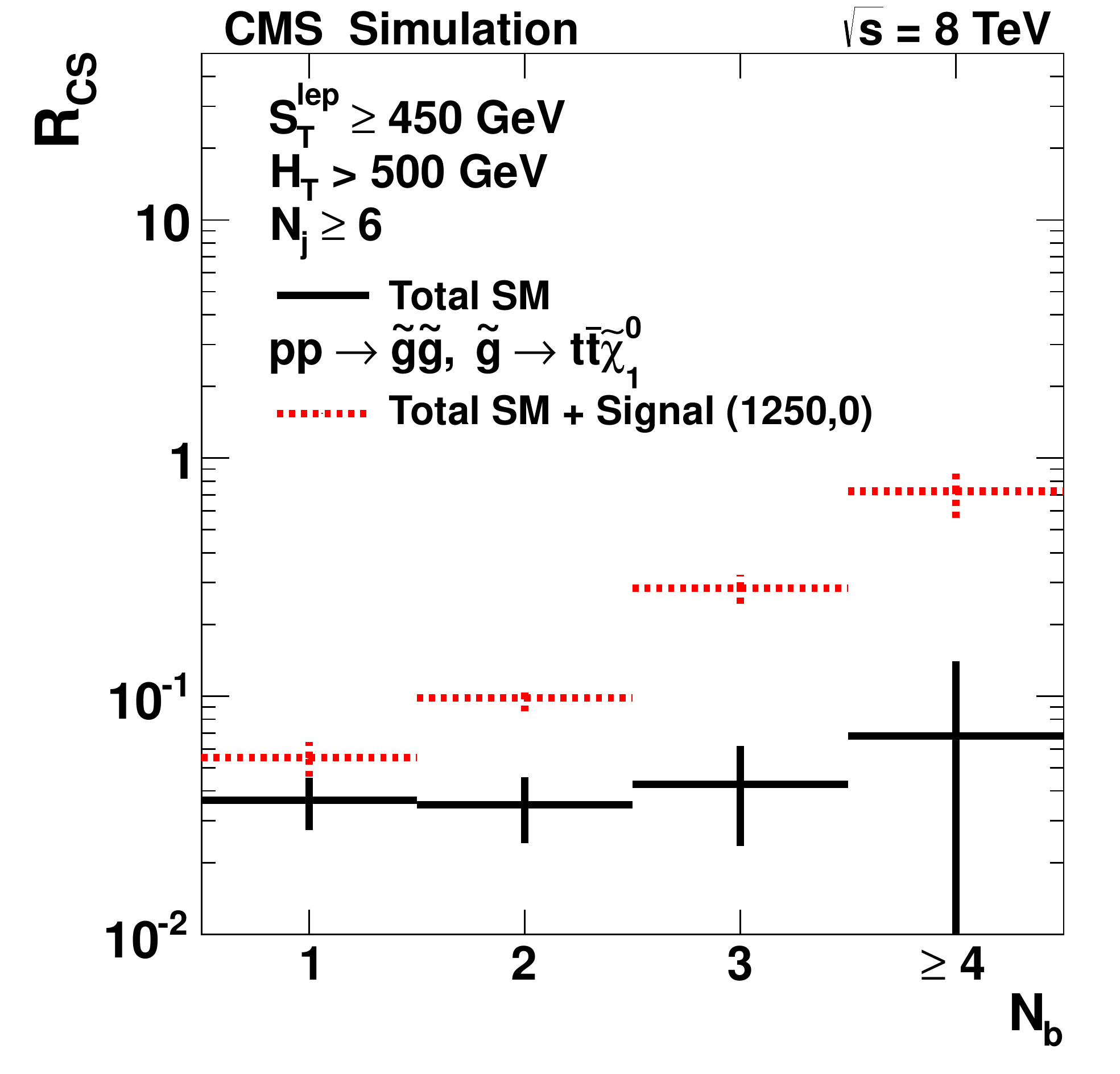}
\caption{The transfer factor \RCSx in simulation for the combined \Pe\ and $\mu$ channels as a function of $\nbjet$ for events with $\njet\ge6$.
The lines correspond to SM only and to the sum of SM backgrounds and signal corresponding to the mass point ($m_{\PSg}=1250\GeV$, $m_{\sTop}=0\GeV$) of model A.
Left: $250 < \stlep < 350\GeV$, center: $350 < \stlep < 450\GeV$, and right: $\stlep > 450\GeV$.
}\label{Rcsplots}
\end{figure*}

The estimate of the SM background in the signal region is obtained using the data and some input from simulation. We define a transfer factor, \RCSx, as the ratio of the number of events with $\Delta\phi(\PW,\ell)> 1$ to the number with $\Delta\phi(\PW,\ell)< 1$.  Figure~\ref{Rcsplots} displays the value of \RCSx as a function of $\nbjet$ for the SM alone and also with the addition of signal from a SUSY benchmark scenario. In the absence of a SUSY signal, the value of \RCSx is roughly independent of the b-jet multiplicity. In the presence of a signal containing four top quarks, however, the value of \RCSx in the $\nbjet \geq 2$ bins changes significantly, whereas it remains unchanged in the $\nbjet = 1$ bin, which is dominated by background from SM processes.
Given this observation, we obtain the transfer factors used to predict the SM background for different values of $\nbjet$, $\RCSpred(\nbjet)$, as $\RCSpred(\nbjet)$=$\RCSx(\nbjet=1)\cdot\kCS(\nbjet)$, where $\RCSx(\nbjet=1)$ is the \RCSx factor measured in data with $\nbjet = 1$ (Table~\ref{tab:RcsNb1Data}) and $\kCS$ is a correction factor obtained from simulation (Table~\ref{tab:default_pred_MC}), introduced to account for any residual dependence of \RCSx on $\nbjet$. The transfer factors $\RCSpred(\nbjet)$ are calculated independently for each bin in $\stlep$.

The calculation of the $\kCS$ factor in simulation is shown in Table~\ref{tab:default_pred_MC}, which lists the yield without a $\kCS$ factor correction, and the observed event yields, as well as the corresponding $\kCS$ correction factors for $\nbjet \ge 3$.
The $\kCS$ factor ranges from 0.93 to 1.45 with statistical uncertainties up to $\pm$0.6.
The large statistical uncertainty reflects the very small event yields expected in the signal region from SM processes.

\begin{table}[b!th]
\topcaption{Data yields and the corresponding \RCSx values for events with $\njet\geq6$ and $\nbjet=1$.}
\label{tab:RcsNb1Data}
\centering
\begin{tabular}{c|c|c|c|c|c} \hline
&&\stlep [\GeVns{}] & Control & Signal   & \RCSx \\ \hline  \hline

\multirow{6}{*}{\begin{sideways} {$\nbjet$=1}  \end{sideways}}
&\multirow{3}{*}{\begin{sideways} \Pe \end{sideways}}
&[250, 350]  & 169 & 6 & 0.04 $\pm$ 0.01\\
&&[350, 450] &  44 & 3 & 0.07 $\pm$ 0.04 \\
&& $>$450   &  17 & 0 & $<$0.06 \\
\cline{2-6}

&\multirow{3}{*}{\begin{sideways} $\mu$ \end{sideways}}
&[250, 350]   & 192 & 9 & 0.05 $\pm$ 0.02\\
&&[350, 450]  &  55 & 2 & 0.04 $\pm$ 0.03\\
&&$>$450     &  10 & 0 & $<$0.1 \\
\hline
\hline
\end{tabular}
\end{table}

\begin{table*}[t!hb]
\centering
\topcaption{Comparison of the simulated yields, combined for the \Pe\ and $\mu$ channels, in the signal region
and the estimate using \RCSx from the $\nbjet=1$ sample.
The \kCS factor is calculated as the ratio of the true yield in the signal region (``MC truth'') and the predicted number.
}
\label{tab:default_pred_MC}
\begin{tabular}{c|c|c|c|c} \hline
&\stlep [\GeVns{}] & Predicted (without $\kCS$) & MC truth  & $\kCS$ \\ \hline  \hline
\multirow{3}{*}{\begin{sideways} $\nbjet$=2 \end{sideways}}
&[250, 350]   & 15.26 $\pm$1.06   & 14.17 $\pm$0.91  & 0.93 $\pm$0.09\\
&[350, 450]  &   2.10 $\pm$0.35   &   3.04 $\pm$0.35 & 1.45 $\pm$0.29\\
&$>$450     &   0.90 $\pm$0.23  &   0.87 $\pm$0.26 & 0.97  $\pm$0.39\\
\hline
\multirow{3}{*}{\begin{sideways} $\nbjet\ge 3$  \end{sideways}}
&[250, 350]   & 2.59 $\pm$0.21   &  3.34$\pm$0.44  &  1.29$\pm$0.20 \\
&[350, 450]  &  0.44 $\pm$ 0.08  &  0.64 $\pm$0.17 &  1.45$\pm$0.47 \\
&$>$450     &  0.18 $\pm$0.06  &  0.22 $\pm$0.09 &   1.22$\pm$0.61 \\
\hline
\end{tabular}
\end{table*}

We observe only a weak dependence of the transfer factor \RCSx on \njet and, as stated above, on \nbjet.
Two sources of this dependence have been identified:
the relative composition of SM samples (\PW+jets, \ttbar($1\ell$), \ttbar$(\ell\ell)$, single top quark),
and the residual dependence of \RCSx within each SM sample.
The effect of each source on \RCSx is found to be $<$50\%. The application of the \kCS factor compensates for these effects and incorporates their uncertainties.
A potential signal would result in much larger values of \RCSx
(e.g., of up to a factor of five larger for the benchmark points) than the variations above, as can be seen from Fig.~\ref{Rcsplots}.

The only elements of the background estimate that depend on simulation are the $\kCS$ factors.
Most potential sources of systematic uncertainties leave $\kCS$ unaffected, since the correction factor reflects only residual changes in the value of \RCSx from $\nbjet=1$ to $\nbjet\geq 3$ ($\nbjet=2$) as a result of each systematic uncertainty.
Systematic uncertainties are estimated as in Section~\ref{sec:METHT}, i.e. by calculating the change induced in the scale factor, \kCS, from various effects and propagating this change to the predicted yields.
The jet\,/\met energy scale and the \cPqb-tagging efficiencies are varied within their uncertainties.
For each independent source (energy scale, heavy- and light-parton tagging efficiencies) the effects of the upwards and downwards variations are averaged.
The \Wjets cross section is varied by 30\% as in Ref.~\cite{SUS-12-010paper}.
The cross section for  \PW+\bbbar is varied by 100\% ~\cite{Zbbmeasure, Wbbmeasure}
and that for single-top-quark production by 50\%~\cite{singletopmeasure}.
We assign an uncertainty of 5 and 10\%, respectively,
 to the W boson and \ttbar polarizations~\cite{WPOLmeasurement,ref:ttbarpolarization}.
These effects are negligible.

Since the estimate of the background in the signal region is based on ratios of events in the data and the $\kCS$ factor that only depends on the number of \cPqb-tagged jets,
the systematic uncertainties of the background prediction
 are expected to be the same for the electron and muon samples.
This is confirmed with an explicit calculation of these uncertainties, and thus the final result uses the combination of the uncertainties from the two lepton flavors.
The overall systematic uncertainty found for $\kCS$, which is dominated by the limited statistics in the simulated samples, is 23\%, 45\% and 70\%, respectively, in the three \stlep ranges.
The total systematic uncertainty of the background
 prediction is dominated by the statistical uncertainty that
 arises due to the limited number of events in the data
 control samples.

\subsection{Multijet background estimate}

Contributions of multijet events to the control and signal regions could affect the correction factors.
Therefore we estimate these contributions from data.
For the muon channel, the MC prediction for the multijet background is smaller than all other backgrounds by two to three orders of magnitude.
This was confirmed by an estimate from data in the previous
single-lepton SUSY search~\cite{SUS-12-010paper}.

In the electron channel, the multijet background is larger than in the muon
channel, but it remains significantly smaller than the other backgrounds.
We make use of the method described in Ref.~\cite{WPOLmeasurement}, employing a control sample in data that is enriched in electrons from multijet events, obtained by inverting some of the electron identification requirements ({\it ``antiselected''} sample).
While the method works well at low \nbjet and \njet, it yields
statistically limited results in the samples with higher \nbjet and
higher \njet.  To obtain more precise predictions for the multijet background in these regions, the estimate from
the $\nbjet=1$ sample is extrapolated with two methods that rely on the relative insensitivity of the multijet background to \nbjet.
The results of these methods are found to be consistent,
and the fraction of multijet events is determined to be less than
5--7\% of the total number of data events observed in the control region.
Based on the antiselected sample, the corresponding transfer factor for multijet events is estimated to be smaller than approximately 2\%.
The multijet contamination in the signal region ($\Delta\phi(\PW,\ell)>1$) is therefore determined to be negligible
and so the multijet background is subtracted only in the control region.

\subsection{Results for signal regions in \texorpdfstring{\stlep}{STlep} and \texorpdfstring{\nbjet}{Nb}}

The background prediction method is validated with
the $3 \le \njet\le 5$ control sample,
which is background dominated with dilepton \ttbar events and with a relative contribution from \Wjets larger than in the signal region.
The compatibility between the predicted and observed yields in this sample is demonstrated by the results shown in the left portion of Table~\ref{tab:FinalResults}.

The predicted and observed data yields in the signal regions are also presented in Table~\ref{tab:FinalResults}.
In the single case of a control region with zero observed events the uncertainty is estimated assuming that one event was present.
Combining all signal bins we predict 19.2$\pm$4.0 events and observe 26.
In the $\nbjet\geq3$ bins, which are the most relevant regions for the signal,
we predict $5.3 \pm 1.5$ events and observe 4.
For $\stlep>350\GeV$ we predict $5.6 \pm 2.5$ events and observe 4.

\begin{table*}[tbh]
\topcaption{
Event yields in data for the $3 \le \njet \le 5$ (validation) and $\njet\geq6$ (signal) samples.
The number of events in the control regions used for the predictions are also shown.
For the lower jet multiplicity validation test, only the statistical uncertainties stemming from
the event counts in the control regions are given,
while statistical and systematic uncertainties are listed for the signal region prediction.
}
\label{tab:FinalResults}
\centering
\begin{tabular}{c|c|c||c|c|c||c|c|c} \hline
&&& \multicolumn{3}{c||}{$3 \le \njet \le 5$} & \multicolumn{3}{c}{$\njet \ge 6$} \\ \hline
&&\stlep [GeV] & Control  & Pred.   & Obs. & Control  & Pred.   & Obs. \\ \hline  \hline
\multirow{6}{*}{\begin{sideways} {$\nbjet=2$}  \end{sideways}}
&\multirow{3}{*}{\begin{sideways} \Pe \end{sideways}}
 & [250, 350] & 548 & 34.2$\pm$5.4 & 30 & 112 & 3.8$\pm$1.8$\pm$0.6 & 9 \\
 && [350, 450] & 174 & 5.1$\pm$1.9 & 8 & 28 & 2.7$\pm$1.9$\pm$0.8 & 2 \\
 && $>$450 & 61 & 5.6$\pm$2.1 & 1 & 9 & 0.0$\pm$0.4$\pm$0.2 & 0 \\ \cline{2-9}
&\multirow{3}{*}{\begin{sideways} $\mu$ \end{sideways}}
 & [250, 350] & 632 & 41.9$\pm$5.6 & 59 & 141 & 6.0$\pm$2.2$\pm$0.9 & 9 \\
 && [350, 450] & 188 & 8.5$\pm$2.4 & 11 & 24 & 1.4$\pm$1.1$\pm$0.4 & 2 \\
 && $>$450 & 71 & 2.5$\pm$1.3 & 1 & 9 & 0.0$\pm$0.7$\pm$0.2 & 0 \\ \hline
\multirow{6}{*}{\begin{sideways} {$\nbjet\ge 3$}  \end{sideways}}
&\multirow{3}{*}{\begin{sideways} \Pe \end{sideways}}
 & [250, 350] & 70 & 3.9$\pm$0.9 & 2 & 45 & 1.9$\pm$0.9$\pm$0.4 & 4 \\
 && [350, 450] & 12 & 0.3$\pm$0.2 & 2 & 7 & 0.9$\pm$0.7$\pm$0.4 & 0 \\
 && $>$450 & 4 & 0.3$\pm$0.2 & 0 & 0 & 0.0$\pm$0.1$\pm$0.03 & 0 \\ \cline{2-9}
&\multirow{3}{*}{\begin{sideways} $\mu$ \end{sideways}}
 & [250, 350] & 59 & 3.9$\pm$0.8 & 5 & 28 & 1.9$\pm$0.8$\pm$0.4 & 0 \\
 && [350, 450] & 25 & 1.1$\pm$0.4 & 0 & 13 & 0.6$\pm$0.5$\pm$0.3 & 0 \\
 && $>$450 & 7 & 0.3$\pm$0.2 & 0 & 2 & 0.0$\pm$0.2$\pm$ 0.1 & 0 \\  \hline
\end{tabular}
\end{table*}

\section{Interpretation}\label{sec:Interpretation}

The compatibility between the observed and predicted event counts in the searches described above
is used to exclude regions in the parameter space of the three models of gluino-mediated production of final states with four top quarks and two LSPs introduced in Section~\ref{sec:Selection}.
The expected signal yield obtained from simulation is corrected for small differences in the efficiencies between data and simulation and for an overestimation of events with high-\pt radiated jets in \MADGRAPH, as described in Ref.~\cite{Chatrchyan:2013xna}.
Systematic uncertainties in the signal yield due to uncertainty in the jet/$\met$ scale~\cite{CMS-PAPERS-JME-10-011}, initial-state radiation, PDFs~\cite{pdf4lhc}, pileup, \cPqb-tagging scale factors~\cite{Chatrchyan:2012jua}, lepton efficiency, and trigger efficiency are calculated for each of the models and for every mass combination.
The uncertainty due to the measurement of the integrated luminosity is 2.6\%~\cite{CMS-PAS-LUM-13-001}.
For model A, the total uncertainty in the signal yields ranges from 20\% to 60\%.
The largest uncertainties are related to the PDFs and occur in regions with small mass differences $m_{\PSg}-m_{\chiz}$ and high $m_{\PSg}$.

The modified-frequentist CL$_{\rm S}$ method~\cite{frequentist_limit,Read:2002hq,LHC-HCG} with a one-sided profile likelihood ratio test statistic is used to define 95\% confidence level (CL) upper limits on the production cross section for each model and mass combination.
Statistical uncertainties related to the observed number of events in control regions are modeled as Poisson distributions.
All other uncertainties are assumed to be multiplicative and are modeled with lognormal distributions.

For each method, several of the signal regions defined in Sections~\ref{sec:METHT} and \ref{sec:DPhi}  are used simultaneously.
In the LS method, three different sets of signal regions are defined,
with a lower \HT bound of either 500, 750, or 1000\GeV.
For each model point, the signal region set with the most stringent expected sensitivity is chosen and the six (\met, \nbjet) bins with $\met > 250\GeV$ are used simultaneously.
The most stringent limits are typically obtained for the lowest \HT threshold.
In the MT method, the requirement $\HT > 750\GeV$ globally yields the best results when combined with the region $400 < \HT < 750\GeV$ in the $\nbjet \ge 3$ bin.
The samples in the two \cPqb-jet multiplicity bins are further divided into \met bins with lower bounds at 250, 350, and 450\GeV.
For the $\nbjet \ge 3$ bin, a low \met region of 150--250\GeV is added, and the \met bins above 250\GeV are combined for $400 < \HT < 750\GeV$.
In the $\Delta\phi(\PW,\ell)$ method all 12 signal regions defined by the three \stlep bins, the two \cPqb-jet multiplicity requirements, and the two lepton flavors are used simultaneously for all model points.
For all three methods, correlations between the uncertainties in different signal regions and between signal yields and background predictions, are taken into account, as well as the effect of signal contamination on the predictions.

\begin{figure*}
\centering 
\includegraphics[height=0.31\textwidth]{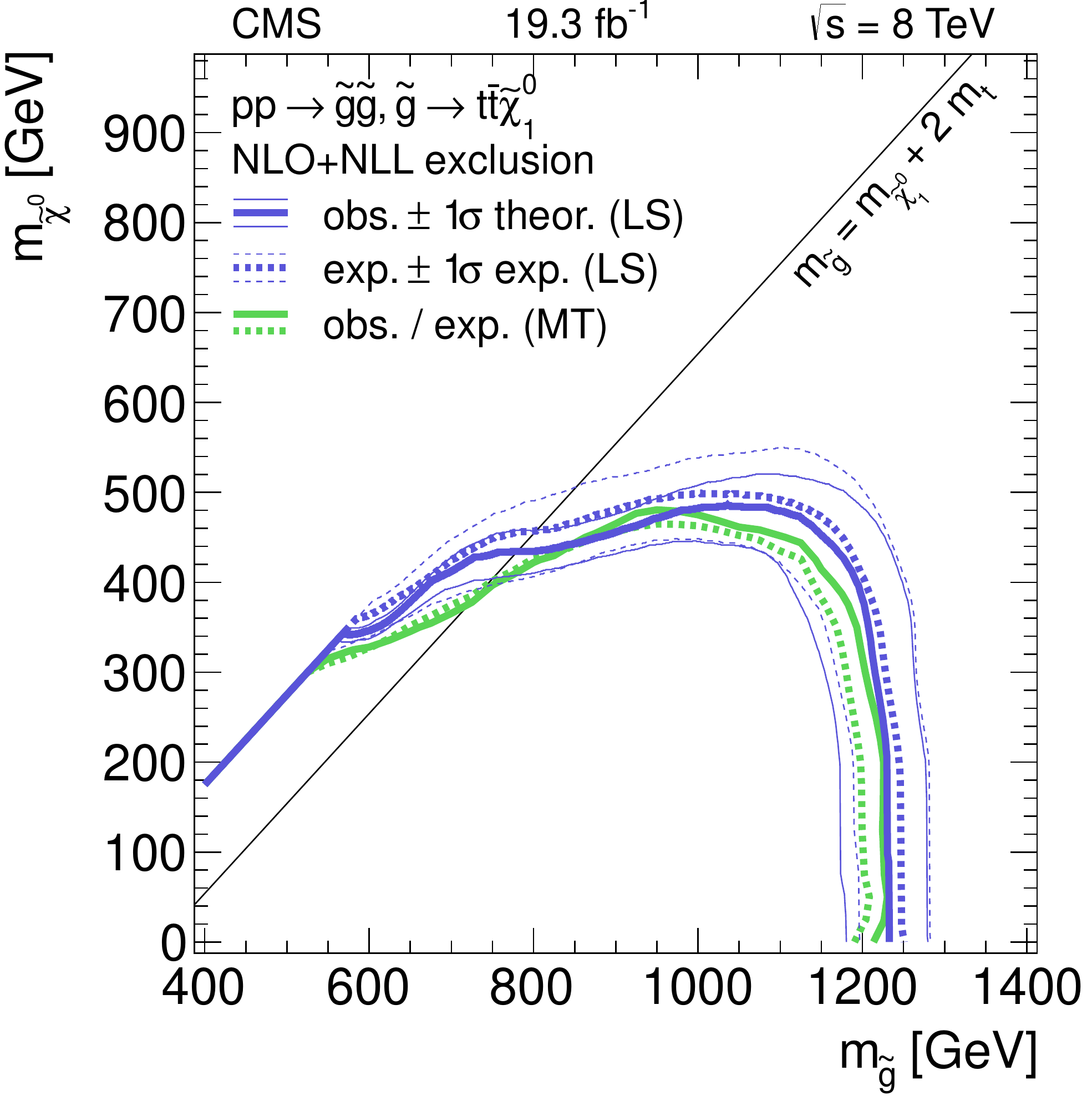}
\includegraphics[height=0.31\textwidth]{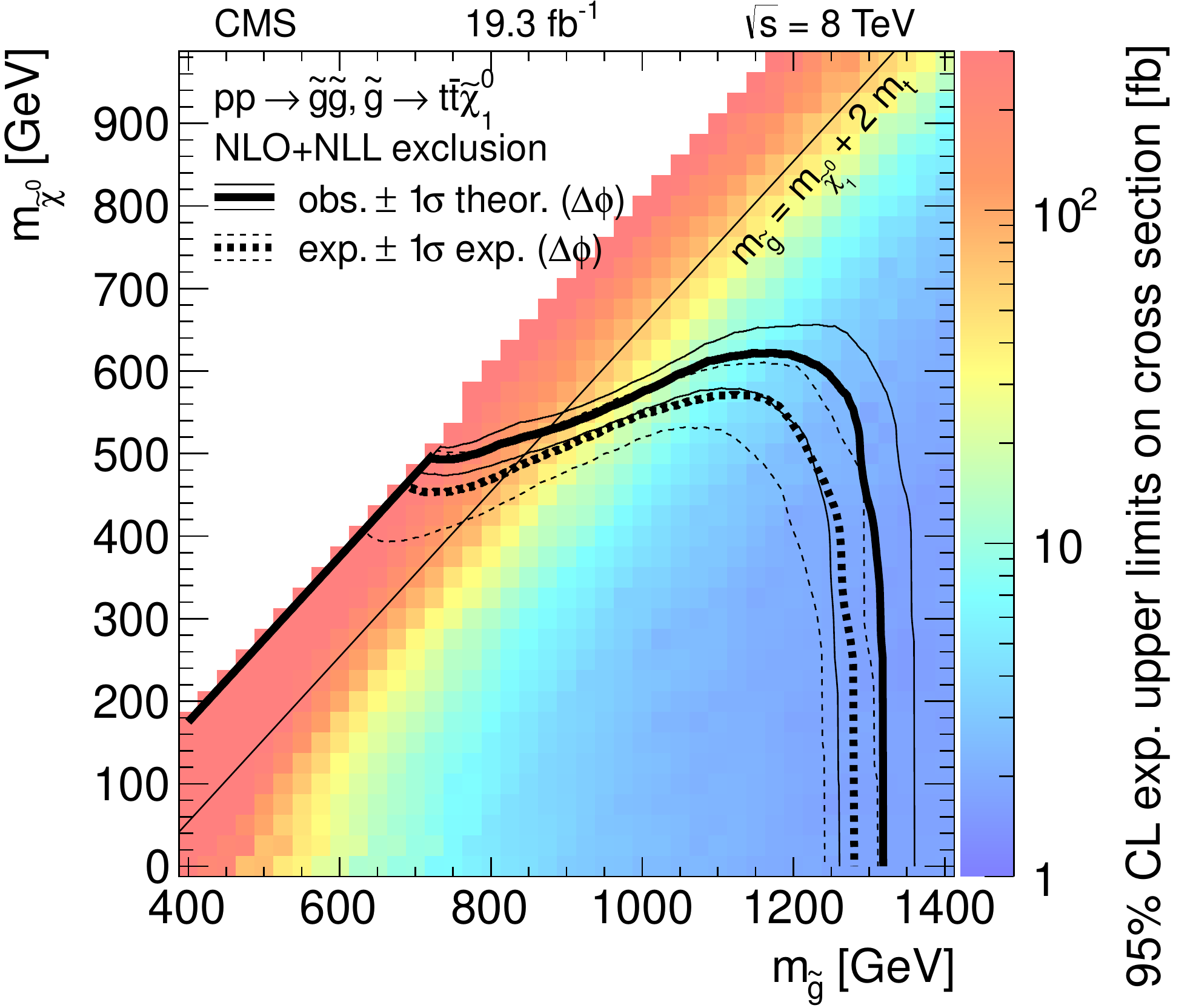}
\includegraphics[height=0.31\textwidth]{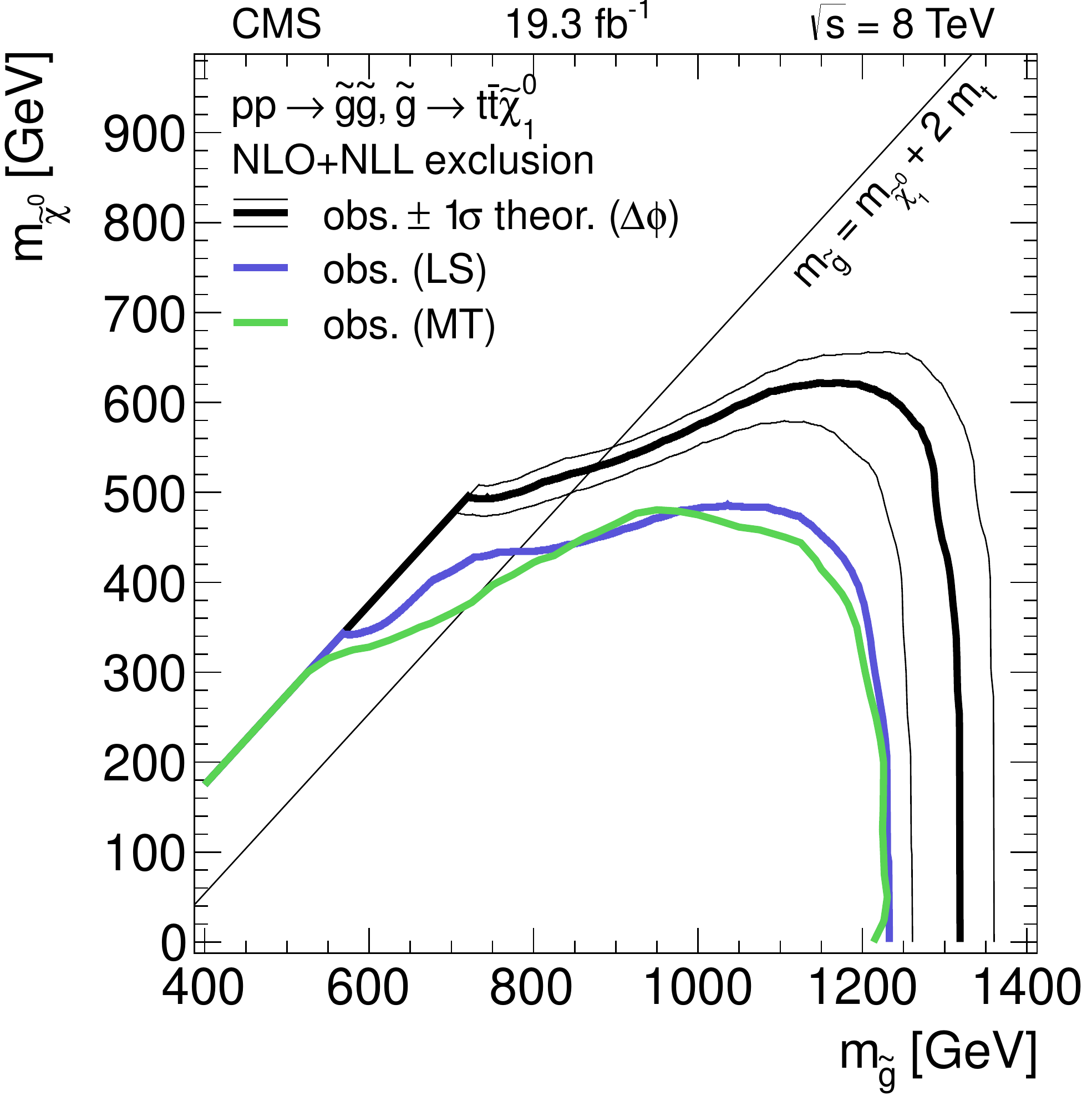}
\includegraphics[height=0.31\textwidth]{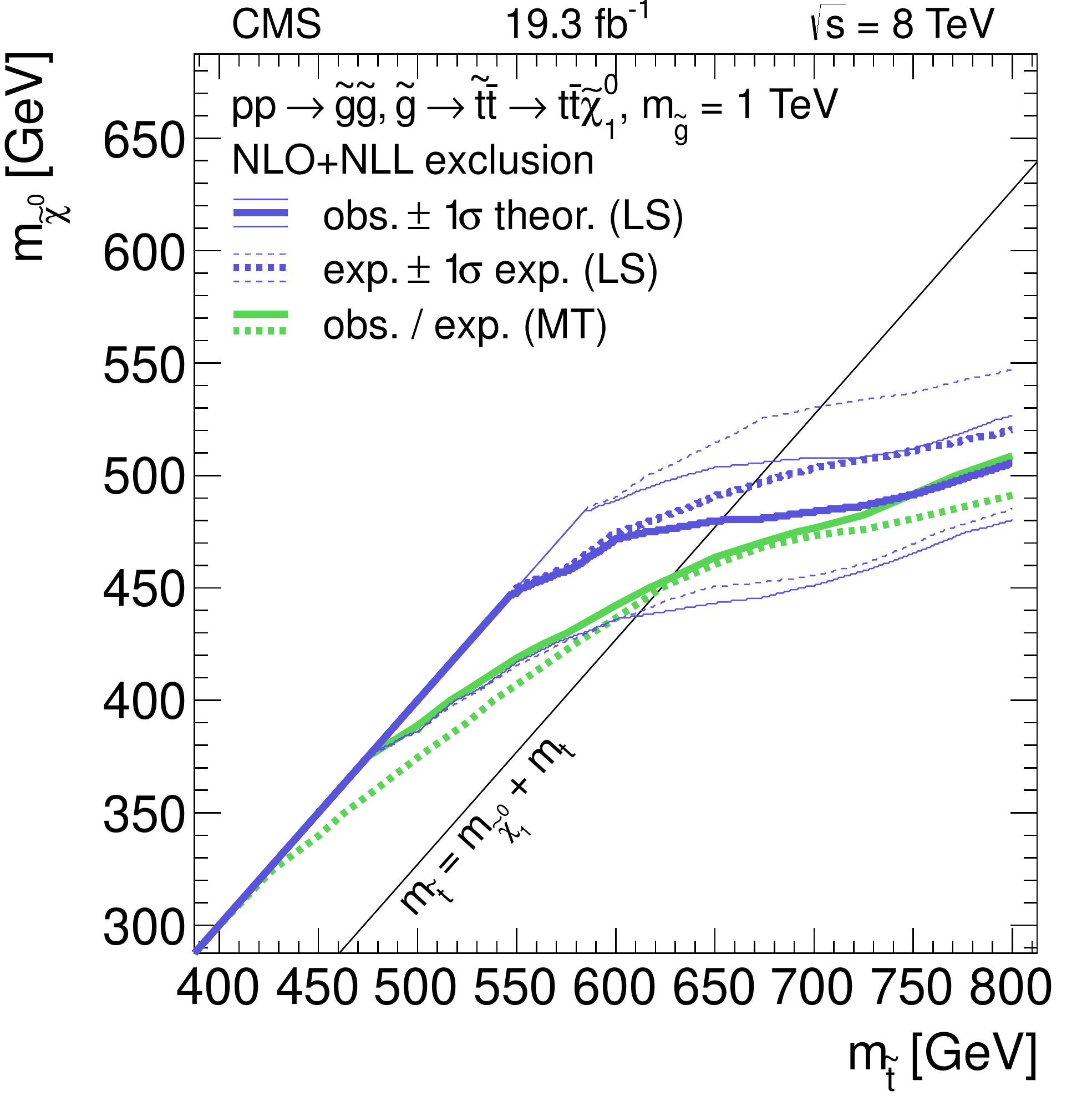}
\includegraphics[height=0.31\textwidth]{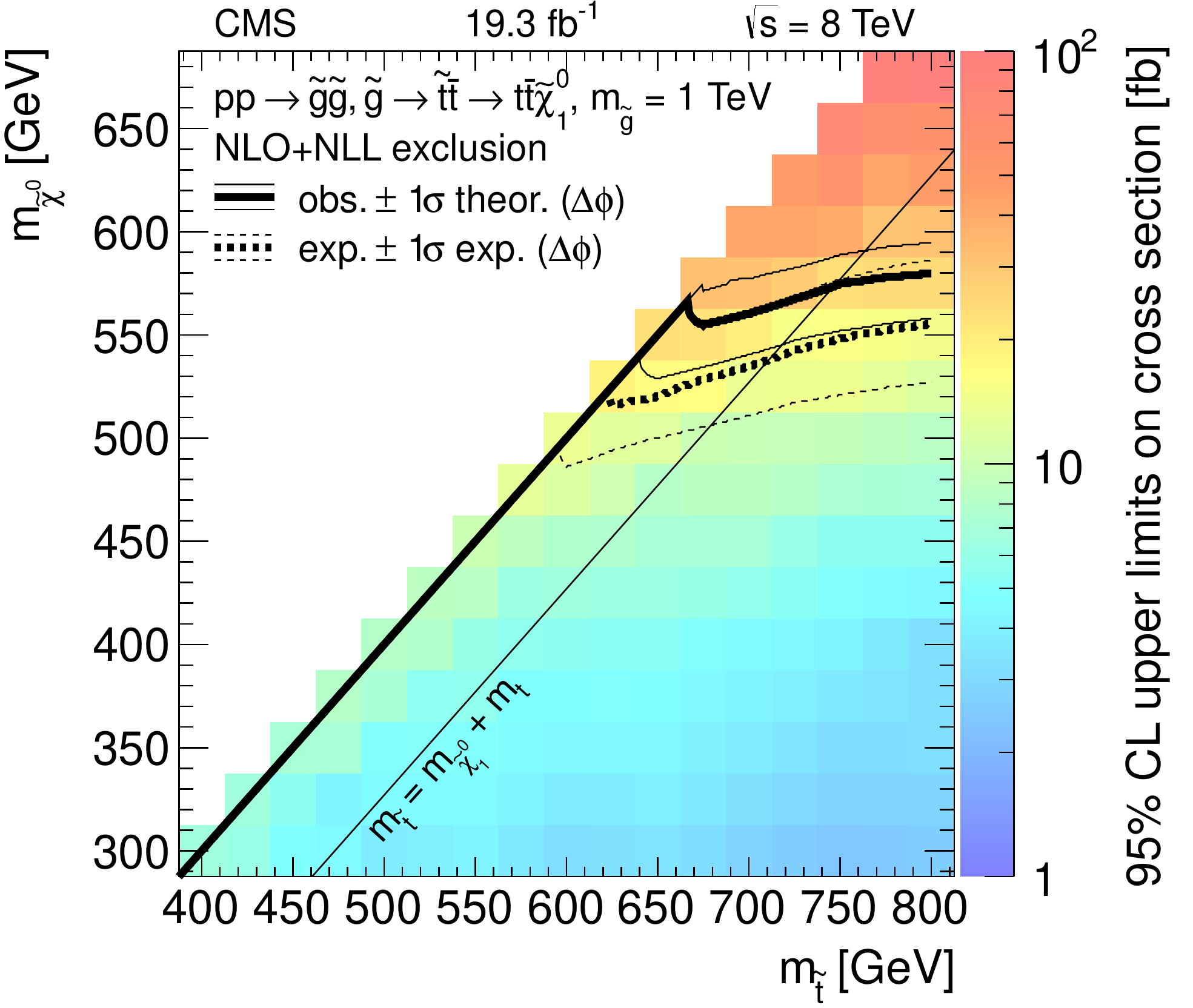}
\includegraphics[height=0.31\textwidth]{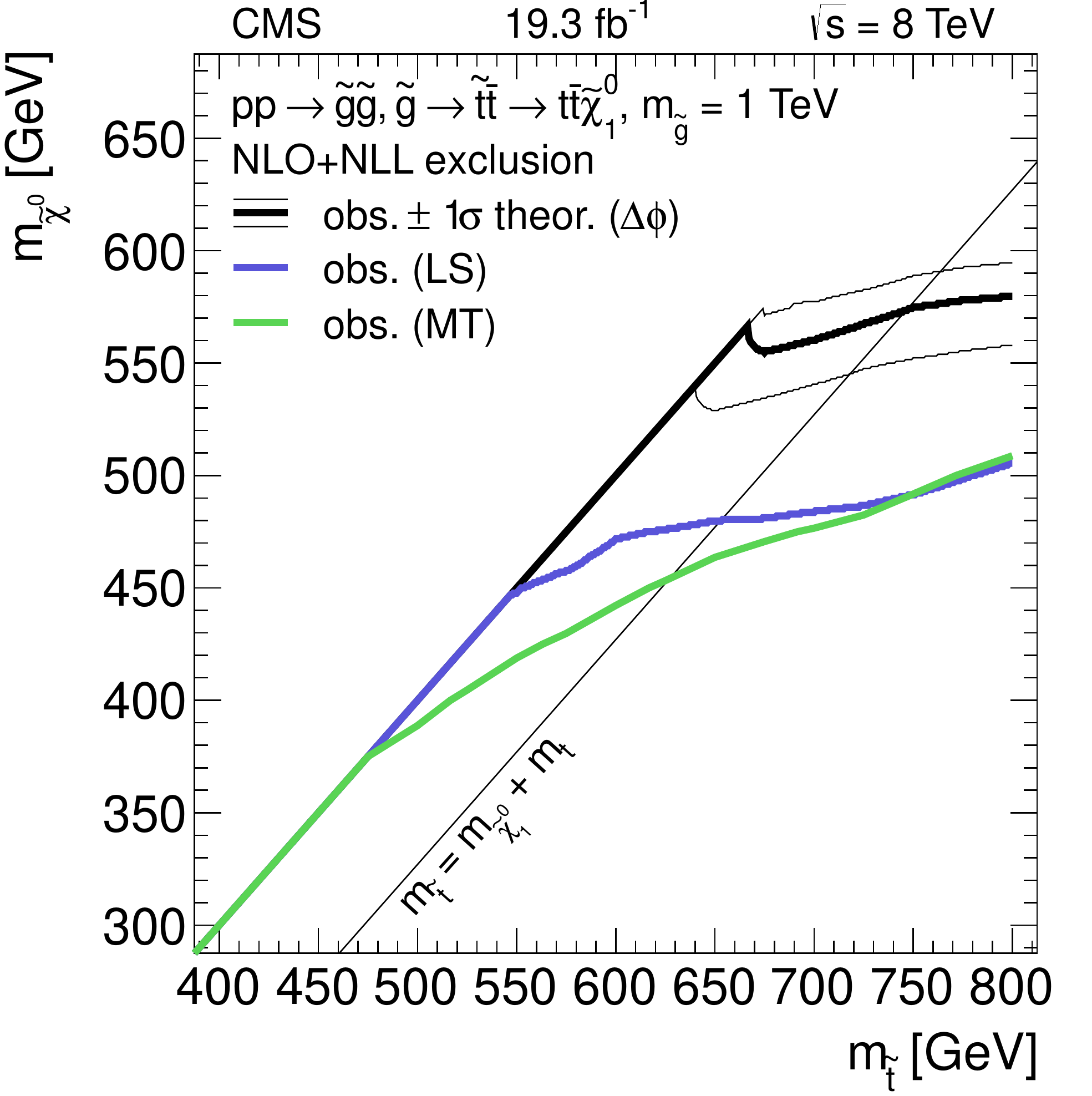}
\includegraphics[height=0.31\textwidth]{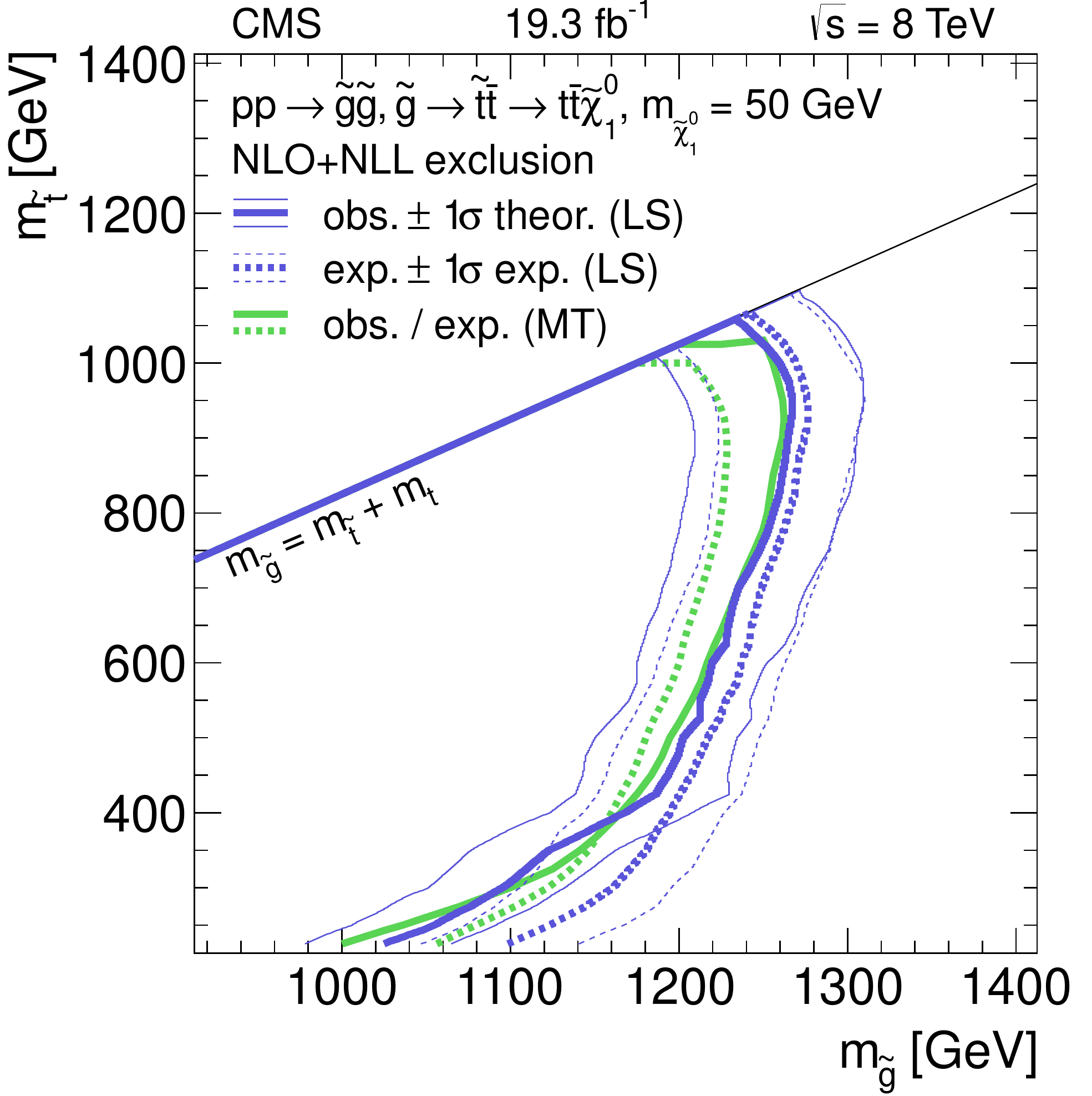}
\includegraphics[height=0.31\textwidth]{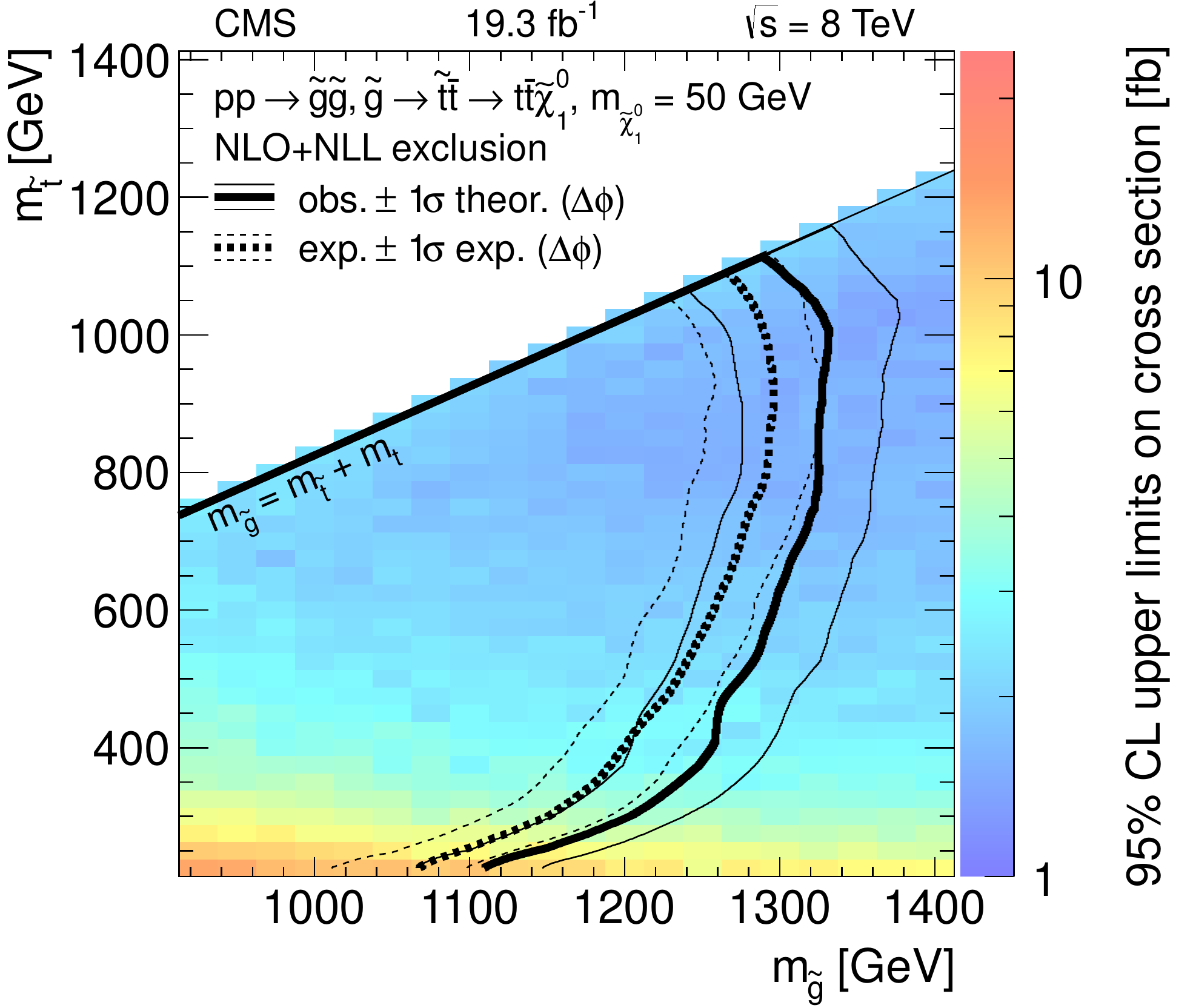}
\includegraphics[height=0.31\textwidth]{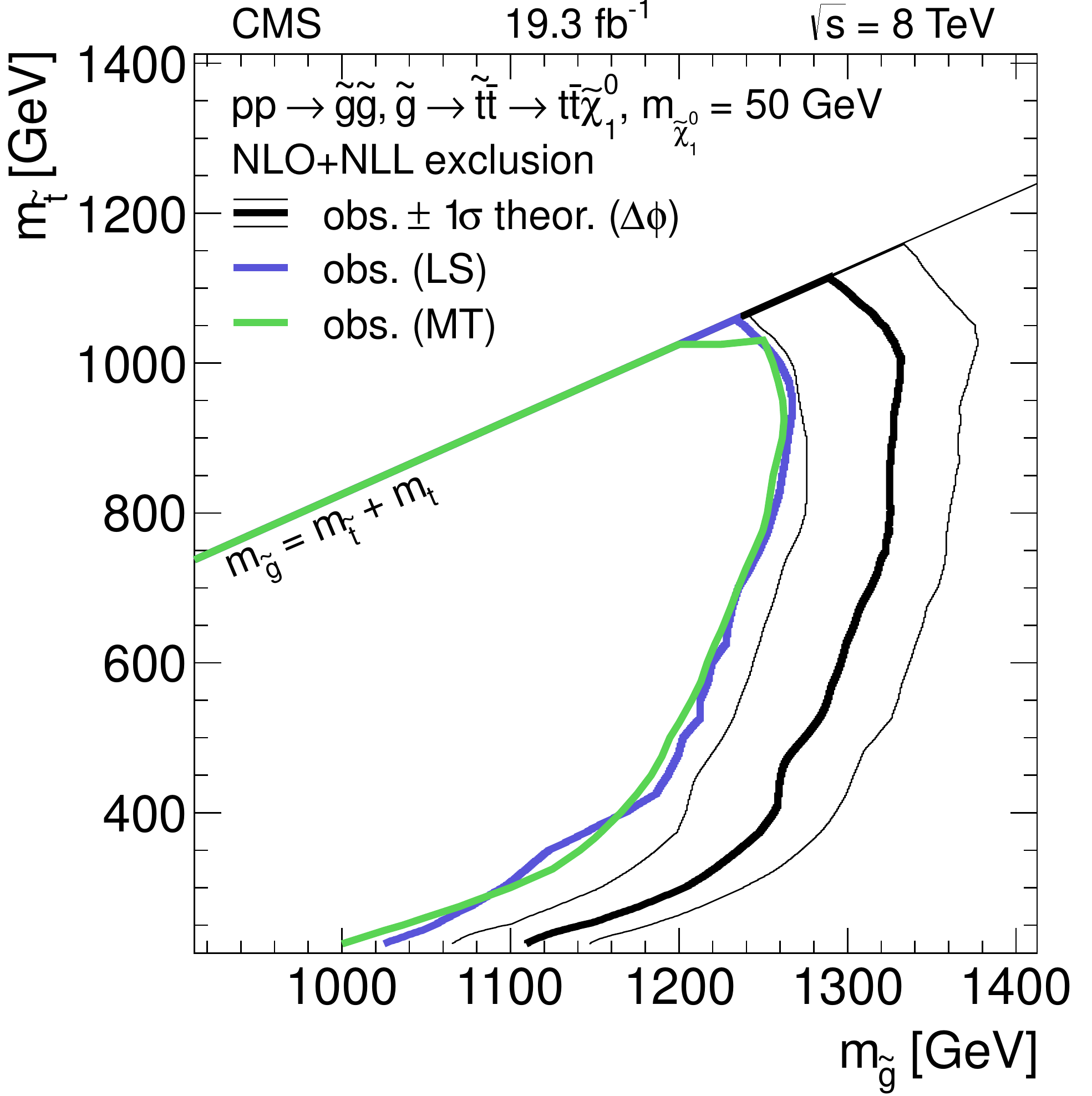}

\caption{
Cross section and mass limits at 95\% CL in the parameter planes of (top) model A, (center) model B, and (bottom) model C.
The color shading indicates the observed limit on the cross section.
The solid (dashed) lines show the observed (expected) mass limits, with the thick lines representing the central value and the thin lines the variations from the theoretical~\cite{Kramer:2012bx} (experimental) uncertainties.
Left column: mass limits for signal regions in \HT and \met (LS and MT) and uncertainty bands for the LS method (the uncertainties for the MT method have similar size).
Central column: cross section and mass limits for signal regions in \stlep and \Dphi (\DP).
Right column: comparison of the observed mass limits for the three methods.
}\label{fig:CombinedLimits}
\end{figure*}

Upper limits on the cross section at a 95\% CL are set in the parameter plane of the three models.
Corresponding mass limits are derived with the next-to-leading order (NLO) + next-to-leading logarithm (NLL) gluino production cross section~\cite{Beenakker:1996ch,PhysRevLett.102.111802,PhysRevD.80.095004,1126-6708-2009-12-041,doi:10.1142/S0217751X11053560} as a reference.
The uncertainty on this cross section is determined as described in Ref.~\cite{Kramer:2012bx}.
These limits are summarized in Fig.~\ref{fig:CombinedLimits}, which shows a comparison of the mass limits obtained for signal regions in \HT and \met, cross section and mass limits for the signal regions in \stlep and \Dphi, and a comparison of the observed mass limits obtained by the three methods.
For each of the considered models the LS and MT methods show a similar reach; the most stringent limits are set by the \DP method.
For model A, with off-shell top squarks, the limits extend to a gluino mass of 1.26\TeV
for the lowest LSP masses and to an LSP mass of 580\GeV for $m_{\PSg}=1.1\TeV$.
At low gluino masses the sensitivity extends to the region $m_{\chiz} > m_{\PSg} - 2 m_{\cPqt}$.
For model B, where the top squarks are on-shell, the limits for $m_{\chiz}$ reach 560\GeV for $m_{\sTop} = 800\GeV$.
For model C the gluino mass limits for low LSP mass are similar to model A for $m_{\sTop} > 500\GeV$ but decrease to $m_{\PSg}=1.0\TeV$
for lower stop masses because the signal populates the lower \met region, which has higher background.
For $m_{\PSg}=1.0\TeV$, the limits cover the full range of top-squark masses if the LSP mass lies below approximately 530\GeV.
Conservatively, these limits are derived from the reference cross section minus one standard deviation~\cite{Kramer:2012bx}.

\section{Summary}\label{sec:Summary}

A sample of proton-proton collisions recorded with the CMS detector at a center-of-mass energy of 8\TeV and corresponding to an integrated luminosity of 19.3\fbinv
has been used for a search for new physics in events with a single isolated electron or muon, multiple high-\pt jets,
including identified \cPqb\ jets, and missing transverse momentum.
This event topology is a possible signature for the production of supersymmetric particles in R-parity conserving models,
in particular the production of gluinos with subsequent decays into top squarks.
The dominant standard model background in a search region defined by the presence of at least six jets,
including at least two jets identified as originating from the fragmentation of \cPqb\ quarks, is due to \ttbar production.

The search is performed with two sets of signal regions, and uses three different methods, each based on data, to estimate the leading background contributions.
The \textit{lepton spectrum} and the  \textit{missing transverse momentum template} methods are designed as searches in the high \HT, high \met region.
They estimate the SM backgrounds (dominated by single-lepton \ttbar decays) for events with two identified \cPqb\ jets and extrapolate these predictions to additional signal regions requiring ${\ge}3$ \cPqb-tagged jets.
The first of these methods uses the lepton \pt distribution to estimate the \met\ spectrum while the second obtains the predictions in a parametrized form by fitting a \ETmiss model to control regions in data.
The \textit{delta phi} method uses the azimuthal angle between the lepton and \PW\ boson directions as a discriminating variable, leading to a strong suppression of the single-lepton backgrounds and leaving dilepton \ttbar events as the leading SM contribution.
The signal regions are defined by the use of the same two \cPqb-jet multiplicity requirements and by bins in \stlep, which probes the total leptonic ($\ell$ and $\nu$) scalar transverse momentum in the event.
While the \textit{delta phi} approach shows the highest sensitivity, the use of different methods, which probe complementary kinematic aspects and both hadronic and leptonic event characteristics, increases the robustness of this search.
Together these methods examine the event sample in both high- and low-yield regions to provide sensitivity to signal topologies with high hadronic activity, missing transverse momentum, and at least two \cPqb\ jets.

No significant excess is observed in any of the signal regions.
Upper limits are set at 95\% CL on the product of production cross section  and branching fraction for three benchmark models of gluino pair production with subsequent decay into virtual or on-shell top squarks, where each of the two top squarks decays in turn into a top quark and the lightest supersymmetric particle.
In the case of decays via virtual top squarks and for light LSPs, gluino masses below 1.26\TeV are excluded.

\section*{Acknowledgments}
We congratulate our colleagues in the CERN accelerator departments for the excellent performance of the LHC and thank the technical and administrative staffs at CERN and at other CMS institutes for their contributions to the success of the CMS effort. In addition, we gratefully acknowledge the computing centres and personnel of the Worldwide LHC Computing Grid for delivering so effectively the computing infrastructure essential to our analyses. Finally, we acknowledge the enduring support for the construction and operation of the LHC and the CMS detector provided by the following funding agencies: BMWF and FWF (Austria); FNRS and FWO (Belgium); CNPq, CAPES, FAPERJ, and FAPESP (Brazil); MES (Bulgaria); CERN; CAS, MoST, and NSFC (China); COLCIENCIAS (Colombia); MSES (Croatia); RPF (Cyprus); MoER, SF0690030s09 and ERDF (Estonia); Academy of Finland, MEC, and HIP (Finland); CEA and CNRS/IN2P3 (France); BMBF, DFG, and HGF (Germany); GSRT (Greece); OTKA and NIH (Hungary); DAE and DST (India); IPM (Iran); SFI (Ireland); INFN (Italy); NRF and WCU (Republic of Korea); LAS (Lithuania); CINVESTAV, CONACYT, SEP, and UASLP-FAI (Mexico); MBIE (New Zealand); PAEC (Pakistan); MSHE and NSC (Poland); FCT (Portugal); JINR (Dubna); MON, RosAtom, RAS and RFBR (Russia); MESTD (Serbia); SEIDI and CPAN (Spain); Swiss Funding Agencies (Switzerland); NSC (Taipei); ThEPCenter, IPST, STAR and NSTDA (Thailand); TUBITAK and TAEK (Turkey); NASU (Ukraine); STFC (United Kingdom); DOE and NSF (USA).
\bibliography{auto_generated}   

\cleardoublepage \appendix\section{The CMS Collaboration \label{app:collab}}\begin{sloppypar}\hyphenpenalty=5000\widowpenalty=500\clubpenalty=5000\input{SUS-13-007-authorlist.tex}\end{sloppypar}
\end{document}

%% file: SUS-13-007-authorlist.tex
\textbf{Yerevan Physics Institute,  Yerevan,  Armenia}\\*[0pt]
S.~Chatrchyan, V.~Khachatryan, A.M.~Sirunyan, A.~Tumasyan
\vskip\cmsinstskip
\textbf{Institut f\"{u}r Hochenergiephysik der OeAW,  Wien,  Austria}\\*[0pt]
W.~Adam, T.~Bergauer, M.~Dragicevic, J.~Er\"{o}, C.~Fabjan\cmsAuthorMark{1}, M.~Friedl, R.~Fr\"{u}hwirth\cmsAuthorMark{1}, V.M.~Ghete, N.~H\"{o}rmann, J.~Hrubec, M.~Jeitler\cmsAuthorMark{1}, W.~Kiesenhofer, V.~Kn\"{u}nz, M.~Krammer\cmsAuthorMark{1}, I.~Kr\"{a}tschmer, D.~Liko, I.~Mikulec, D.~Rabady\cmsAuthorMark{2}, B.~Rahbaran, H.~Rohringer, R.~Sch\"{o}fbeck, J.~Strauss, A.~Taurok, W.~Treberer-Treberspurg, W.~Waltenberger, C.-E.~Wulz\cmsAuthorMark{1}
% ...
\vskip\cmsinstskip
% ...
1:~~Also at Vienna University of Technology, Vienna, Austria\\
2:~~Also at CERN, European Organization for Nuclear Research, Geneva, Switzerland\\
% ...


%% file: SUS-13-007_temp.bbl
\providecommand{\href}[2]{#2}\begingroup\raggedright\begin{thebibliography}{10}%
\makeatletter
\providecommand{\hrefCMSnoop }[0]{\@secondoftwo}%
\makeatother
\providecommand{\doi}{\texttt{doi:}\begingroup \urlstyle{tt}\Url}

\bibitem{Wess:1974tw}
\hrefCMSnoop {} {J.~Wess and B.~Zumino, ``{Supergauge Transformations in
  Four-Dimensions}'',} \textit{ Nucl. Phys. B} \textbf{ 70} (1974) 39,
\href{http://dx.doi.org/10.1016/0550-3213(74)90355-1}{\doi{10.1016/0550-3213(74)90355-1}}.

\bibitem{Dimopoulos:1981zb}
\hrefCMSnoop {} {S.~Dimopoulos and H.~Georgi, ``{Softly Broken Supersymmetry
  and SU(5)}'',} \textit{ Nucl. Phys. B} \textbf{ 193} (1981) 150,
\href{http://dx.doi.org/10.1016/0550-3213(81)90522-8}{\doi{10.1016/0550-3213(81)90522-8}}.

\bibitem{Nilles:1983ge}
\hrefCMSnoop {} {H.~P. Nilles, ``{Supersymmetry, Supergravity and Particle
  Physics}'',} \textit{ Phys. Rept.} \textbf{ 110} (1984) 1,
\href{http://dx.doi.org/10.1016/0370-1573(84)90008-5}{\doi{10.1016/0370-1573(84)90008-5}}.

\bibitem{Haber:1984rc}
\hrefCMSnoop {} {H.~E. Haber and G.~L. Kane, ``{The Search for Supersymmetry:
  Probing Physics Beyond the Standard Model}'',} \textit{ Phys. Rept.} \textbf{
  117} (1985) 75,
\href{http://dx.doi.org/10.1016/0370-1573(85)90051-1}{\doi{10.1016/0370-1573(85)90051-1}}.

\bibitem{Barbieri:1982eh}
\hrefCMSnoop {} {R.~Barbieri, S.~Ferrara, and C.~A. Savoy, ``{Gauge models with
  spontaneously broken local supersymmetry}'',} \textit{ Phys. Lett. B}
  \textbf{ 119} (1982) 343,
\href{http://dx.doi.org/10.1016/0370-2693(82)90685-2}{\doi{10.1016/0370-2693(82)90685-2}}.

\bibitem{Dawson:1983fw}
\hrefCMSnoop {} {S.~Dawson, E.~Eichten, and C.~Quigg, ``{Search for
  Supersymmetric Particles in Hadron--Hadron Collisions}'',} \textit{ Phys.
  Rev. D} \textbf{ 31} (1985) 1581,
\href{http://dx.doi.org/10.1103/PhysRevD.31.1581}{\doi{10.1103/PhysRevD.31.1581}}.

\bibitem{Sakai:1981gr}
\hrefCMSnoop {} {N.~Sakai, ``{Naturalness in supersymmetric GUTS}'',} \textit{
  Z. Phys. C} \textbf{ 11} (1981) 153,
\href{http://dx.doi.org/10.1007/BF01573998}{\doi{10.1007/BF01573998}}.

\bibitem{Dimopoulos:1995mi}
\hrefCMSnoop {} {S.~Dimopoulos and G.~F. Giudice, ``{Naturalness constraints in
  supersymmetric theories with nonuniversal soft terms}'',} \textit{ Phys.
  Lett. B} \textbf{ 357} (1995) 573,
  \href{http://dx.doi.org/10.1016/0370-2693(95)00961-J}{\doi{10.1016/0370-2693(95)00961-J}},
\href{http://www.arXiv.org/abs/hep-ph/9507282}{\texttt{ arXiv:hep-ph/9507282}}.

\bibitem{Papucci:2011wy}
\hrefCMSnoop {} {M.~Papucci, J.~T. Ruderman, and A.~Weiler, ``{Natural SUSY
  endures}'',} \textit{ JHEP} \textbf{ 09} (2012) 035,
  \href{http://dx.doi.org/10.1007/JHEP09(2012)035}{\doi{10.1007/JHEP09(2012)035}},
\href{http://www.arXiv.org/abs/1110.6926}{\texttt{ arXiv:1110.6926}}.

\bibitem{Brust:2011tb}
\hrefCMSnoop {} {C.~Brust, A.~Katz, S.~Lawrence, and R.~Sundrum, ``{SUSY, the
  Third Generation and the LHC}'',} \textit{ JHEP} \textbf{ 03} (2012) 103,
  \href{http://dx.doi.org/10.1007/JHEP03(2012)103}{\doi{10.1007/JHEP03(2012)103}},
\href{http://www.arXiv.org/abs/1110.6670}{\texttt{ arXiv:1110.6670}}.

\bibitem{Chatrchyan:2013xna}
\hrefCMSnoop {} {{ CMS} Collaboration, ``{Search for top-squark pair production
  in the single-lepton final state in pp collisions at {$\sqrt{s} = 8
  \TeV$}}'',} \textit{ Eur. Phys. J. C} \textbf{ 73} (2013) 2677,
  \href{http://dx.doi.org/10.1140/epjc/s10052-013-2677-2}{\doi{10.1140/epjc/s10052-013-2677-2}},
\href{http://www.arXiv.org/abs/1308.1586}{\texttt{ arXiv:1308.1586}}.

\bibitem{SUS-12-010paper}
\hrefCMSnoop {} {{ CMS} Collaboration, ``{Search for supersymmetry in {\Pp\Pp}\
  collisions at {$\sqrt{s}=7 \TeV$} in events with a single lepton, jets, and
  missing transverse momentum}'',} \textit{ Eur. Phys. J. C} \textbf{ 73}
  (2013) 2404,
  \href{http://dx.doi.org/10.1140/epjc/s10052-013-2404-z}{\doi{10.1140/epjc/s10052-013-2404-z}},
\href{http://www.arXiv.org/abs/1212.6428}{\texttt{ arXiv:1212.6428}}.

\bibitem{SUS-11-028paper}
\hrefCMSnoop {} {{ CMS} Collaboration, ``{Search for supersymmetry in final
  states with a single lepton, {\cPqb}-quark jets, and missing transverse
  energy in proton-proton collisions at {$\sqrt{s} = 7 \TeV$}}'',} \textit{
  Phys. Rev. D} \textbf{ 87} (2013) 052006,
  \href{http://dx.doi.org/10.1103/PhysRevD.87.052006}{\doi{10.1103/PhysRevD.87.052006}},
  \href{http://www.arXiv.org/abs/1211.3143}{\texttt{ arXiv:1211.3143}}.

\bibitem{Aad:2012naa}
\hrefCMSnoop {} {{ ATLAS} Collaboration, ``{Multi-channel search for squarks
  and gluinos in {$\sqrt{s}=7 \TeV$} {\Pp\Pp}\ collisions with the ATLAS
  detector}'',} \textit{ Eur. Phys. J. C} \textbf{ 73} (2013) 2362,
  \href{http://dx.doi.org/10.1140/epjc/s10052-013-2362-5}{\doi{10.1140/epjc/s10052-013-2362-5}},
\href{http://www.arXiv.org/abs/1212.6149}{\texttt{ arXiv:1212.6149}}.

\bibitem{Aad:2012yr}
\hrefCMSnoop {} {{ ATLAS} Collaboration, ``{Search for light top squark pair
  production in final states with leptons and {\cPqb}-jets with the ATLAS
  detector in {$\sqrt{s}=7 \TeV$} proton-proton collisions}'',} \textit{ Phys.
  Lett. B} \textbf{ 720} (2013) 13,
  \href{http://dx.doi.org/10.1016/j.physletb.2013.01.049}{\doi{10.1016/j.physletb.2013.01.049}},
\href{http://www.arXiv.org/abs/1209.2102}{\texttt{ arXiv:1209.2102}}.

\bibitem{Aad:2012ms}
\hrefCMSnoop {} {{ ATLAS} Collaboration, ``{Further search for supersymmetry at
  {$\sqrt{s}=7 \TeV$} in final states with jets, missing transverse momentum
  and isolated leptons with the ATLAS detector}'',} \textit{ Phys. Rev. D}
  \textbf{ 86} (2012) 092002,
  \href{http://dx.doi.org/10.1103/PhysRevD.86.092002}{\doi{10.1103/PhysRevD.86.092002}},
\href{http://www.arXiv.org/abs/1208.4688}{\texttt{ arXiv:1208.4688}}.

\bibitem{:2012ar}
\hrefCMSnoop {} {{ ATLAS} Collaboration, ``{Search for direct top squark pair
  production in final states with one isolated lepton, jets, and missing
  transverse momentum in {$\sqrt{s}=7 \TeV$} {\Pp\Pp}\ collisions using 4.7
  {\fbinv} of ATLAS data}'',} \textit{ Phys. Rev. Lett.} \textbf{ 109} (2012)
  211803,
  \href{http://dx.doi.org/10.1103/PhysRevLett.109.211803}{\doi{10.1103/PhysRevLett.109.211803}},
\href{http://www.arXiv.org/abs/1208.2590}{\texttt{ arXiv:1208.2590}}.

\bibitem{Farrar:1978xj}
\hrefCMSnoop {} {G.~R. Farrar and P.~Fayet, ``Phenomenology of the production,
  decay, and detection of new hadronic states associated with supersymmetry'',}
  \textit{ Phys. Lett. B} \textbf{ 76} (1978) 575,
\href{http://dx.doi.org/10.1016/0370-2693(78)90858-4}{\doi{10.1016/0370-2693(78)90858-4}}.

\bibitem{ArkaniHamed:2007fw}
N.~Arkani-Hamed\hrefCMSnoop {} { {et~al.}, ``{MARMOSET: The Path from LHC Data
  to the New Standard Model via On-Shell Effective Theories}'',} (2007).
\href{http://www.arXiv.org/abs/hep-ph/0703088}{\texttt{ arXiv:hep-ph/0703088}}.

\bibitem{Alwall:2008ag}
\hrefCMSnoop {} {J.~Alwall, P.~C. Schuster, and N.~Toro, ``{Simplified models
  for a first characterization of new physics at the LHC}'',} \textit{ Phys.
  Rev. D} \textbf{ 79} (2009) 075020,
  \href{http://dx.doi.org/10.1103/PhysRevD.79.075020}{\doi{10.1103/PhysRevD.79.075020}},
\href{http://www.arXiv.org/abs/0810.3921}{\texttt{ arXiv:0810.3921}}.

\bibitem{Alves:2011wf}
\hrefCMSnoop {} {D.~Alves {et~al.}, ``{Simplified models for LHC new physics
  searches}'',} \textit{ J. Phys. G} \textbf{ 39} (2012) 105005,
  \href{http://dx.doi.org/10.1088/0954-3899/39/10/105005}{\doi{10.1088/0954-3899/39/10/105005}},
\href{http://www.arXiv.org/abs/1105.2838}{\texttt{ arXiv:1105.2838}}.

\bibitem{ref:CMS}
\hrefCMSnoop {} {{ CMS} Collaboration, ``The {CMS} experiment at the {CERN
  LHC}'',} \textit{ JINST} \textbf{ 03} (2008) S08004,
  \href{http://dx.doi.org/10.1088/1748-0221/3/08/S08004}{\doi{10.1088/1748-0221/3/08/S08004}}.

\bibitem{madgraph}
J.~Alwall\hrefCMSnoop {} { {et~al.}, ``{MadGraph 5}: going beyond'',} \textit{
  JHEP} \textbf{ 06} (2011) 128,
  \href{http://dx.doi.org/10.1007/JHEP06(2011)128}{\doi{10.1007/JHEP06(2011)128}},
\href{http://www.arXiv.org/abs/1106.0522}{\texttt{ arXiv:1106.0522}}.

\bibitem{Pumplin:2002vw}
J.~Pumplin\hrefCMSnoop {} { {et~al.}, ``{New generation of parton distributions
  with uncertainties from global QCD analysis}'',} \textit{ JHEP} \textbf{ 07}
  (2002) 012,
  \href{http://dx.doi.org/10.1088/1126-6708/2002/07/012}{\doi{10.1088/1126-6708/2002/07/012}},
\href{http://www.arXiv.org/abs/hep-ph/0201195}{\texttt{ arXiv:hep-ph/0201195}}.

\bibitem{powheg}
\hrefCMSnoop {} {S.~Frixione, P.~Nason, and C.~Oleari, ``{Matching NLO QCD
  computations with parton shower simulations: the POWHEG method}'',} \textit{
  JHEP} \textbf{ 11} (2007) 070,
  \href{http://dx.doi.org/10.1088/1126-6708/2007/11/070}{\doi{10.1088/1126-6708/2007/11/070}},
\href{http://www.arXiv.org/abs/0709.2092}{\texttt{ arXiv:0709.2092}}.

\bibitem{pythia}
\hrefCMSnoop {} {T.~Sj{\"o}strand, S.~Mrenna, and P.~Z. Skands, ``{PYTHIA} 6.4
  physics and manual'',} \textit{ JHEP} \textbf{ 05} (2006) 026,
  \href{http://dx.doi.org/10.1088/1126-6708/2006/05/026}{\doi{10.1088/1126-6708/2006/05/026}},
\href{http://www.arXiv.org/abs/hep-ph/0603175}{\texttt{ arXiv:hep-ph/0603175}}.

\bibitem{Chatrchyan:1363299}
\hrefCMSnoop {} {{ CMS} Collaboration, ``{Measurement of the underlying event
  activity at the LHC with {$\sqrt{s} = 7 \TeV$} and comparison with {$\sqrt{s}
  = 0.9 \TeV$}}'',} \textit{ JHEP} \textbf{ 09} (2011) 109,
  \href{http://dx.doi.org/10.1007/JHEP09(2011)109}{\doi{10.1007/JHEP09(2011)109}},
  \href{http://www.arXiv.org/abs/1107.0330}{\texttt{ arXiv:1107.0330}}.

\bibitem{tauola}
\hrefCMSnoop {} {Z.~W{\c{a}}s, ``{TAUOLA the library for tau lepton decay, and
  KKMC/KORALB/KORALZ/\ldots report}'',} \textit{ Nucl. Phys. Proc. Suppl.}
  \textbf{ 98} (2001) 96,
  \href{http://dx.doi.org/10.1016/S0920-5632(01)01200-2}{\doi{10.1016/S0920-5632(01)01200-2}},
  \href{http://www.arXiv.org/abs/hep-ph/0011305}{\texttt{
  arXiv:hep-ph/0011305}}.

\bibitem{GEANT4}
\hrefCMSnoop {} {{ GEANT4} Collaboration, ``{GEANT4}---a simulation toolkit'',}
  \textit{ Nucl. Instrum. Meth. A} \textbf{ 506} (2003) 250,
\href{http://dx.doi.org/10.1016/S0168-9002(03)01368-8}{\doi{10.1016/S0168-9002(03)01368-8}}.

\bibitem{Abdullin:2011zz}
\hrefCMSnoop {} {{ CMS} Collaboration, ``{The fast simulation of the CMS
  detector at LHC}'',} \textit{ J. Phys. Conf. Ser.} \textbf{ 331} (2011)
  032049,
\href{http://dx.doi.org/10.1088/1742-6596/331/3/032049}{\doi{10.1088/1742-6596/331/3/032049}}.

\bibitem{ref:eleID}
\href {http://cdsweb.cern.ch/record/1299116} {{ CMS} Collaboration, ``Electron
  reconstruction and identification at {$\sqrt{s} = 7 \TeV$}'',} CMS Physics
  Analysis Summary CMS-PAS-EGM-10-004, 2010.

\bibitem{CMS-PAPERS-MUO-10-004}
\hrefCMSnoop {} {{ CMS} Collaboration, ``Performance of {CMS} muon
  reconstruction in pp collision events at {$\sqrt{s} = 7 \TeV$}'',} \textit{
  JINST} \textbf{ 7} (2012) P10002,
  \href{http://dx.doi.org/10.1088/1748-0221/7/10/P10002}{\doi{10.1088/1748-0221/7/10/P10002}},
  \href{http://www.arXiv.org/abs/1206.4071}{\texttt{ arXiv:1206.4071}}.

\bibitem{CMS:2011aa}
\hrefCMSnoop {} {{ CMS} Collaboration, ``{Measurement of the inclusive {\PW}\
  and {\cPZ}\ production cross sections in {\Pp\Pp}\ collisions at
  {$\sqrt{s}=7\TeV$}}'',} \textit{ JHEP} \textbf{ 10} (2011) 132,
  \href{http://dx.doi.org/10.1007/JHEP10(2011)132}{\doi{10.1007/JHEP10(2011)132}},
\href{http://www.arXiv.org/abs/1107.4789}{\texttt{ arXiv:1107.4789}}.

\bibitem{ref:PAS-PFT-10-002}
\href {http://cdsweb.cern.ch/record/1279341} {{ CMS} Collaboration,
  ``Commissioning of the Particle-Flow Reconstruction in Minimum-Bias and Jet
  Events from {\Pp\Pp} Collisions at 7 {\TeV}'',} CMS Physics Analysis Summary
  CMS-PAS-PFT-10-002, 2010.

\bibitem{ref:antiKT}
\hrefCMSnoop {} {M.~Cacciari, G.~P. Salam, and G.~Soyez, ``The anti-{$k_t$} jet
  clustering algorithm'',} \textit{ JHEP} \textbf{ 04} (2008) 063,
  \href{http://dx.doi.org/10.1088/1126-6708/2008/04/063}{\doi{10.1088/1126-6708/2008/04/063}},
  \href{http://www.arXiv.org/abs/0802.1189}{\texttt{ arXiv:0802.1189}}.

\bibitem{CMS-PAPERS-JME-10-011}
\hrefCMSnoop {} {{ CMS} Collaboration, ``Determination of jet energy
  calibration and transverse momentum resolution in {CMS}'',} \textit{ JINST}
  \textbf{ 6} (2011) P11002,
  \href{http://dx.doi.org/10.1088/1748-0221/6/11/P11002}{\doi{10.1088/1748-0221/6/11/P11002}},
  \href{http://www.arXiv.org/abs/1107.4277}{\texttt{ arXiv:1107.4277}}.

\bibitem{Chatrchyan:2012jua}
\hrefCMSnoop {} {{ CMS} Collaboration, ``{Identification of {\cPqb}-quark jets
  with the CMS experiment}'',} \textit{ JINST} \textbf{ 8} (2013) P04013,
  \href{http://dx.doi.org/10.1088/1748-0221/8/04/P04013}{\doi{10.1088/1748-0221/8/04/P04013}},
\href{http://www.arXiv.org/abs/1211.4462}{\texttt{ arXiv:1211.4462}}.

\bibitem{CMS-PAS-BTV-13-001}
\href {http://cds.cern.ch/record/1581306} {{ CMS} Collaboration, ``Performance
  of {\cPqb}\ tagging at {$\sqrt{s} = 8 \TeV$} in multijet, {\ttbar} and
  boosted topology events'',} CMS Physics Analysis Summary CMS-PAS-BTV-13-001,
  2013.

\bibitem{ref:leptonspectrum}
\hrefCMSnoop {} {V.~Pavlunin, ``Modeling missing transverse energy in
  {$V$}+jets at {CERN LHC}'',} \textit{ Phys. Rev. D} \textbf{ 81} (2010)
  035005,
  \href{http://dx.doi.org/10.1103/PhysRevD.81.035005}{\doi{10.1103/PhysRevD.81.035005}},
  \href{http://www.arXiv.org/abs/0906.5016}{\texttt{ arXiv:0906.5016}}.

\bibitem{ref:ttbarpolarization}
\hrefCMSnoop {} {A.~Czarnecki, J.~G. K{\"o}rner, and J.~H. Piclum, ``Helicity
  fractions of {\PW}\ bosons from top quark decays at next-to-next-to-leading
  order in {QCD}'',} \textit{ Phys. Rev. D} \textbf{ 81} (2010) 111503(R),
  \href{http://dx.doi.org/10.1103/PhysRevD.81.111503}{\doi{10.1103/PhysRevD.81.111503}},
  \href{http://www.arXiv.org/abs/1005.2625}{\texttt{ arXiv:1005.2625}}.

\bibitem{ref:Pickands}
\hrefCMSnoop {} {J.~Pickands, ``Statistical inference using extreme order
  statistics'',} \textit{ Annals of Stat.} \textbf{ 3} (1975) 119,
  \href{http://dx.doi.org/10.1214/aos/1176343003}{\doi{10.1214/aos/1176343003}}.

\bibitem{Zbbmeasure}
\hrefCMSnoop {} {{ CMS} Collaboration, ``Measurement of the
  {\cPZ}/{$\gamma^*$}+{\cPqb}-jet cross section in pp collisions at {$\sqrt{s}
  = 7\TeV$}'',} \textit{ JHEP} \textbf{ 06} (2012) 126,
  \href{http://dx.doi.org/10.1007/JHEP06(2012)126}{\doi{10.1007/JHEP06(2012)126}},
  \href{http://www.arXiv.org/abs/1204.1643}{\texttt{ arXiv:1204.1643}}.

\bibitem{Wbbmeasure}
\hrefCMSnoop {} {{ ATLAS} Collaboration, ``{Measurement of the cross-section
  for W boson production in association with {\cPqb}-jets in pp collisions at
  {$\sqrt{s} = 7 \TeV$} with the ATLAS detector}'',} \textit{ JHEP} \textbf{
  06} (2013) 084,
  \href{http://dx.doi.org/10.1007/JHEP06(2013)084}{\doi{10.1007/JHEP06(2013)084}},
\href{http://www.arXiv.org/abs/1302.2929}{\texttt{ arXiv:1302.2929}}.

\bibitem{singletopmeasure}
\hrefCMSnoop {} {{ CMS} Collaboration, ``{Measurement of the $t$-channel single
  top quark production cross section in {\Pp\Pp}\ collisions at {$\sqrt{s}=7
  \TeV$}}'',} \textit{ Phys. Rev. Lett.} \textbf{ 107} (2011) 091802,
  \href{http://dx.doi.org/10.1103/PhysRevLett.107.091802}{\doi{10.1103/PhysRevLett.107.091802}},
\href{http://www.arXiv.org/abs/1106.3052}{\texttt{ arXiv:1106.3052}}.

\bibitem{WPOLmeasurement}
\hrefCMSnoop {} {{ CMS} Collaboration, ``Measurement of the polarization of
  {\PW}\ bosons with large transverse momenta in {\PW}+jets events at the
  {LHC}'',} \textit{ Phys. Rev. Lett.} \textbf{ 107} (2011) 021802,
  \href{http://dx.doi.org/10.1103/PhysRevLett.107.021802}{\doi{10.1103/PhysRevLett.107.021802}},
  \href{http://www.arXiv.org/abs/1104.3829}{\texttt{ arXiv:1104.3829}}.

\bibitem{pdf4lhc}
M.~Botje\hrefCMSnoop {} { {et~al.}, ``The {PDF4LHC} {W}orking {G}roup Interim
  Recommendations'',} (2011).
\href{http://www.arXiv.org/abs/1101.0538}{\texttt{ arXiv:1101.0538}}.

\bibitem{CMS-PAS-LUM-13-001}
\href {http://cds.cern.ch/record/1598864} {{ CMS} Collaboration, ``CMS
  Luminosity Based on Pixel Cluster Counting - Summer 2013 Update'',} CMS
  Physics Analysis Summary CMS-PAS-LUM-13-001, 2013.

\bibitem{frequentist_limit}
\hrefCMSnoop {} {T.~Junk, ``{Confidence level computation for combining
  searches with small statistics}'',} \textit{ Nucl. Instrum. Meth. A} \textbf{
  434} (1999) 435,
  \href{http://dx.doi.org/10.1016/S0168-9002(99)00498-2}{\doi{10.1016/S0168-9002(99)00498-2}},
\href{http://www.arXiv.org/abs/hep-ex/9902006}{\texttt{ arXiv:hep-ex/9902006}}.

\bibitem{Read:2002hq}
\hrefCMSnoop {} {A.~L. Read, ``Presentation of search results: the {CL$_{\rm
  S}$} technique'',} \textit{ J. Phys. G} \textbf{ 28} (2002) 2693,
  \href{http://dx.doi.org/10.1088/0954-3899/28/10/313}{\doi{10.1088/0954-3899/28/10/313}}.

\bibitem{LHC-HCG}
\href {http://cdsweb.cern.ch/record/1379837} {{ATLAS and CMS Collaborations,
  LHC Higgs Combination Group}, ``Procedure for the LHC Higgs boson search
  combination in Summer 2011'',} Technical Report ATL-PHYS-PUB/2011-11, CMS
  NOTE 2011/005, 2011.

\bibitem{Beenakker:1996ch}
\hrefCMSnoop {} {W.~Beenakker, R.~H{\"o}pker, M.~Spira, and P.~M. Zerwas,
  ``Squark and gluino production at hadron colliders'',} \textit{ Nucl. Phys.
  B} \textbf{ 492} (1997) 51,
  \href{http://dx.doi.org/10.1016/S0550-3213(97)80027-2}{\doi{10.1016/S0550-3213(97)80027-2}},
  \href{http://www.arXiv.org/abs/hep-ph/9610490}{\texttt{
  arXiv:hep-ph/9610490}}.

\bibitem{PhysRevLett.102.111802}
\hrefCMSnoop {} {A.~Kulesza and L.~Motyka, ``Threshold Resummation for
  Squark-Antisquark and Gluino-Pair Production at the {LHC}'',} \textit{ Phys.
  Rev. Lett.} \textbf{ 102} (2009) 111802,
  \href{http://dx.doi.org/10.1103/PhysRevLett.102.111802}{\doi{10.1103/PhysRevLett.102.111802}},
  \href{http://www.arXiv.org/abs/0807.2405}{\texttt{ arXiv:0807.2405}}.

\bibitem{PhysRevD.80.095004}
\hrefCMSnoop {} {A.~Kulesza and L.~Motyka, ``Soft gluon resummation for the
  production of gluino-gluino and squark-antisquark pairs at the {LHC}'',}
  \textit{ Phys. Rev. D} \textbf{ 80} (2009) 095004,
  \href{http://dx.doi.org/10.1103/PhysRevD.80.095004}{\doi{10.1103/PhysRevD.80.095004}},
  \href{http://www.arXiv.org/abs/0905.4749}{\texttt{ arXiv:0905.4749}}.

\bibitem{1126-6708-2009-12-041}
W.~Beenakker\hrefCMSnoop {} { {et~al.}, ``Soft-gluon resummation for squark and
  gluino hadroproduction'',} \textit{ JHEP} \textbf{ 09} (2009) 041,
  \href{http://dx.doi.org/10.1088/1126-6708/2009/12/041}{\doi{10.1088/1126-6708/2009/12/041}},
  \href{http://www.arXiv.org/abs/0909.4418}{\texttt{ arXiv:0909.4418}}.

\bibitem{doi:10.1142/S0217751X11053560}
W.~Beenakker\hrefCMSnoop {} { {et~al.}, ``Squark and gluino hadroproduction'',}
  \textit{ Int. J. Mod. Phys. A} \textbf{ 26} (2011) 2637,
  \href{http://dx.doi.org/10.1142/S0217751X11053560}{\doi{10.1142/S0217751X11053560}},
  \href{http://www.arXiv.org/abs/1105.1110}{\texttt{ arXiv:1105.1110}}.

\bibitem{Kramer:2012bx}
M.~Kr{\"a}mer\hrefCMSnoop {} { {et~al.}, ``{Supersymmetry production cross
  sections in pp collisions at {$\sqrt{s} = 7 \TeV$}}'',} (2012).
\href{http://www.arXiv.org/abs/1206.2892}{\texttt{ arXiv:1206.2892}}.

\end{thebibliography}\endgroup
